\documentclass{aa}  
\usepackage{natbib}
\usepackage[FIGTOPCAP]{subfigure}
\usepackage{multirow}
\usepackage{txfonts}
\usepackage[dvipsnames]{xcolor}
\usepackage{tablefootnote}
\usepackage{scrextend}
\usepackage{booktabs}

\makeatletter
\def\NAT@spacechar{~}% NEW
\makeatother

\usepackage{hyperref}

\hypersetup{
    colorlinks=true,
    citecolor=RoyalBlue,
    filecolor=black,
    linkcolor=PineGreen,
    urlcolor=RoyalBlue
}

\usepackage{etoolbox}
\makeatletter
\patchcmd\@combinedblfloats{\box\@outputbox}{\unvbox\@outputbox}{}{%
   \errmessage{\noexpand\@combinedblfloats could not be patched}%
}%
 \makeatother

\begin{document} 

\defcitealias{paper1}{Paper I}

   \title{Diffuse interstellar bands in the \ion{H}{ii} region M17}
\subtitle{Insights into their relation with the total-to-selective visual extinction $R_V$}

   \author{M.C. Ram\'irez-Tannus \inst{1}
           \and
          N.L.J. Cox\inst{2}
          \and
          L. Kaper\inst{1}
          \and 
          A. de Koter\inst{1,3}
}

   \institute{Anton Pannekoek Institute for Astronomy, University of Amsterdam,
              Science Park 904, 1098 XH Amsterdam, The Netherlands\\
              \email{m.c.ramireztannus@uva.nl}
        \and
        ACRI-ST, 260 Route du Pin Montard, Sophia Antipolis, France
		\and 
		Institute of Astronomy, 
		KU Leuven, Celestijnenlaan 200D, 3001 Leuven, Belgium
             }

   \date{\today}

 \abstract{Diffuse interstellar bands (DIBs) are broad absorption features measured in sightlines probing the diffuse interstellar medium. Although large carbon-bearing molecules have been proposed as the carriers producing DIBs, their identity remains unknown. DIBs make an important contribution to the extinction curve; the sight line to the young massive star-forming region M17 shows anomalous extinction in the sense that the total-to-selective extinction parameter differs significantly from the average Galactic value and may reach values $R_{V} > 4$. Anomalous DIBs have been reported in the sight line towards Herschel~36 ($R_{V} = 5.5$), in the massive star-forming region M8. Higher values of $R_V$ have been associated with a relatively higher fraction of large dust grains in the line of sight.}
 {Given the high $R_V$ values, we investigate whether the DIBs in sight lines towards young OB stars in M17 show a peculiar behaviour.}
 {We measure the properties of the most prominent DIBs in M17 and study these as a function of $E(B-V)$ and $R_{V}$. We also analyse the gaseous and dust components contributing to the interstellar extinction.} 
 {The DIB strengths in M17 concur with the observed relations between DIB equivalent width and reddening $E(B-V)$ in Galactic sight lines. For several DIBs we discover a linear relation between the normalised DIB strength EW/$A_{V}$ and $R_{V}^{-1}$. 
These trends suggest two groups of DIBs: (i) a group of ten moderately strong DIBs that show a sensitivity to changes in $R_{V}$ that is modest and proportional to DIB strength, and (ii) a group of four very strong DIBs that react sensitively and to a similar degree to changes in $R_{V}$, but in a way that does not appear to depend on DIB strength.}
 {The DIB behaviour as a function of reddening is not peculiar in sight lines to M17. Also, we do not detect anomalous DIB profiles as seen in Herschel~36. DIBs are stronger, per unit visual extinction, in sight lines characterised by a smaller value of $R_{V}$, i.e.\ those sight lines that contain a relatively large fraction of small dust particles. New relations between extinction normalised DIB strengths, EW/$A_V$, and $R_V$ support the idea that DIB carriers and interstellar dust are intimately connected. Furthermore, given the distinct behaviour of two groups of DIBs, different types of carriers do not necessarily relate to the dust grains in a similar way.}


   \maketitle
%
%________________________________________________________________

\section{Introduction}
\label{P3:sec:intro}

The nature of the carrier(s) of diffuse interstellar bands (DIBs) is one of the oldest mysteries in stellar spectroscopy. Over 400 DIBs have been observed in the optical wavelength range \citep[e.g.][]{2009ApJ...705...32H} and about a dozen DIBs have been detected in the near infrared \citep{2011Natur.479..200G, 2014A&A...569A.117C, 2015ApJ...800..137H}. DIBs are thought to be large carbon-bearing molecules and may represent the largest reservoir of organic material in the Universe \citep{2014IAUS..297....3S}. Laboratory experiments simulating interstellar conditions recently proposed C$_{60}^{+}$ as the carrier of the $\lambda$9577 and $\lambda$9632 DIBs \citep{2015Natur.523..322C}, but the vast majority of the DIBs remains unidentified. 
For a recent overview of DIBs, see \citet{2014IAUS..297.....C}.

DIBs measured in sight lines towards massive star forming regions are reported to behave differently compared to the average Galactic sight line. \citet{1997ApJ...489..698H} noted that the DIBs observed in the direction of M17, over the extinction range of $A_V =$ 3 - 10, show little change in spectral shape nor a significant increase in strength. They suggested that either the DIB features are already saturated at a small value of the visual extinction $A_{V}$, or that the interstellar material local to M17, where the increased extinction is being traced, does not contain the DIB carriers. \citet{2013ApJ...773...41D} and \citet{2013ApJ...773...42O} detected anomalously broad DIBs at $\lambda\lambda$5780.5, 5797.1, and 6613.6 in the sight line to Herschel~36, an O star multiple system associated with the Hourglass nebula in the giant H~{\sc ii} region M8 (Lagoon Nebula). The DIBs show an excess of absorption in the red wing of the profile; excited lines of CH and CH$^{+}$ are detected as well. \citet{2013ApJ...773...42O} interpret this observation as being caused by infrared pumping of rotational levels of relatively small molecules. 

Sight lines towards massive star forming regions (Orion Trapezium region, Carina, M8) often show a high value of the total-to-selective extinction parameter $R_{V}$ \citep[see e.g.][]{1989ApJ...345..245C, 2007ApJ...663..320F}. The common interpretation is that the value of $R_{V}$ characterises the dust particle size distribution. Sight lines that include relatively many small dust particles display an extinction curve with a small value of $R_{V}$ and produce stronger interstellar absorption at short (UV) wavelengths, and vice versa. Most Galactic sight lines have $R_{V} \sim 3.1 \pm 0.3$ \citep[][]{2007ApJ...663..320F} and we refer to extinction curves with higher or lower values of $R_{V}$ as anomalous.

The strongest feature in the extinction curve is the 2175~\AA\  bump; although the carrier of this feature also remains unidentified, it is generally attributed to carbonaceous particles, either in the form of graphite, a mixture of hydrogenated amorphous carbon grains and polycyclyc aromatic hydrocarbons (PAHs), or various aromatic forms of carbon \citep[e.g.][]{1977ApJ...217..425M, 1998ApJ...507L.177M, 2015ApJ...809..120M}. It is believed that the presence and strength of the bump depends on the metallicity of the environment as it appears slightly weaker in the LMC extinction curves, and it is essentially absent in the SMC extinction curve \citep{1998ApJ...500..816G, 1999ApJ...515..128M}. 
Some sightlines towards the LMC (specifically the LMC\,2 supershell near 30 Dor) and the SMC sight lines have small $R_{V}$ values \citep[$\sim 2.7$,][]{2003ApJ...594..279G}. 
Remarkably, the one SMC line of sight that exhibits normal strength DIBs (and CH) shows Galactic-type dust extinction and includes the 2175~\AA\ extinction feature \citep{2002ApJ...576L.117E,2006ApJS..165..138W, 2007A&A...470..941C}. 
Furthermore, Galactic sight lines that are SMC-like show very weak DIBs per unit reddening (\mbox{\citealt{2002ApJ...573..670S}}; \mbox{\citealt{2007A&A...470..941C}}).

If the carriers of the DIBs and those of the 2175~\AA\ bump are produced from the same initial dust through the same physical process, or if the DIB carriers originate from the fragmentation of the 2175~\AA\ bump carriers, one would expect them to be related. 
More recently \citet{2011ApJ...733...91X} studied the relation between DIBs and the 2175~\AA\ bump in detail. They collected 2175~\AA\ bump and DIB strength measurements, from the literature, towards 84 interstellar sightlines for eight DIBs and found no significant correlation between the two. They explain the lack of a correlation by hypothesising that DIB carriers correspond to the smallest, free-flying PAH molecules and ions, while the 2175~\AA\ carriers correspond to the larger or clustered PAHs.

In \citet[][hereafter \citetalias{paper1}]{paper1} we studied a sample of young massive (pre-)main-sequence stars in the giant \ion{H}{ii} region M17 located at a distance of $1.98_{-0.12}^{+0.14}$~kpc \citep{2011ApJ...733...25X}. We obtained optical to near-infrared VLT/X-shooter spectra and derived the extinction parameters by modelling the spectral energy distribution. The sight lines are characterised by anomalous extinction ($R_{V}$ in the range 3.3 to 4.6 and $A_{V}$ between 5 and 14~mag). M17 is one of the most luminous star-forming regions in the Galaxy with a luminosity if $3.6\times10^{6}$~L$_\sun$ \citep{2007A&A...462..123P}. It contains about 16 O stars and more than 100 B stars; its age is $\leq 1$~Myr \citep[][\citetalias{paper1}]{1997ApJ...489..698H,2008ApJ...686..310H,2007ApJS..169..353B,2009ApJ...696.1278P}.

In this paper we present a detailed analysis of the DIBs in eight sight lines to M17, taking into account the anomalous  
extinction observed in this region ($R_V > 3.1$). Earlier work hinted at the peculiar behaviour of the DIBs in these sight lines \citep{1997ApJ...489..698H} and we investigate whether the DIB properties are somehow related to the extinction caused by dust. Such a relation may shed new light on the physical and chemical nature of the DIB crriers.

The paper is organised as follows: in the next section we briefly describe the data set and reduction procedure, and in Section~\ref{P3:sec:extinction} we characterise the interstellar extinction in the direction of M17, both the gaseous and the dust component. In Section~\ref{P3:sec:DIBs} the DIB properties are presented. Subsequently, we compare the DIB properties to those observed in other Galactic sight lines (Section~\ref{P3:sec:OtherSightlines}). In Section~\ref{P3:sec:DIBvsRv} we address the observed dependence of the DIBs in M17 on the value of $R_{V}$. In Section~\ref{P3:sec:discussion} we discuss the results in the context of the anomalous extinction and the properties of the interstellar dust. In the last section we summarise our conclusions.

\section{X-shooter observations}
\label{P3:sec:obs}

  \begin{figure}[t!]
   \centering
   \includegraphics[width=\hsize]{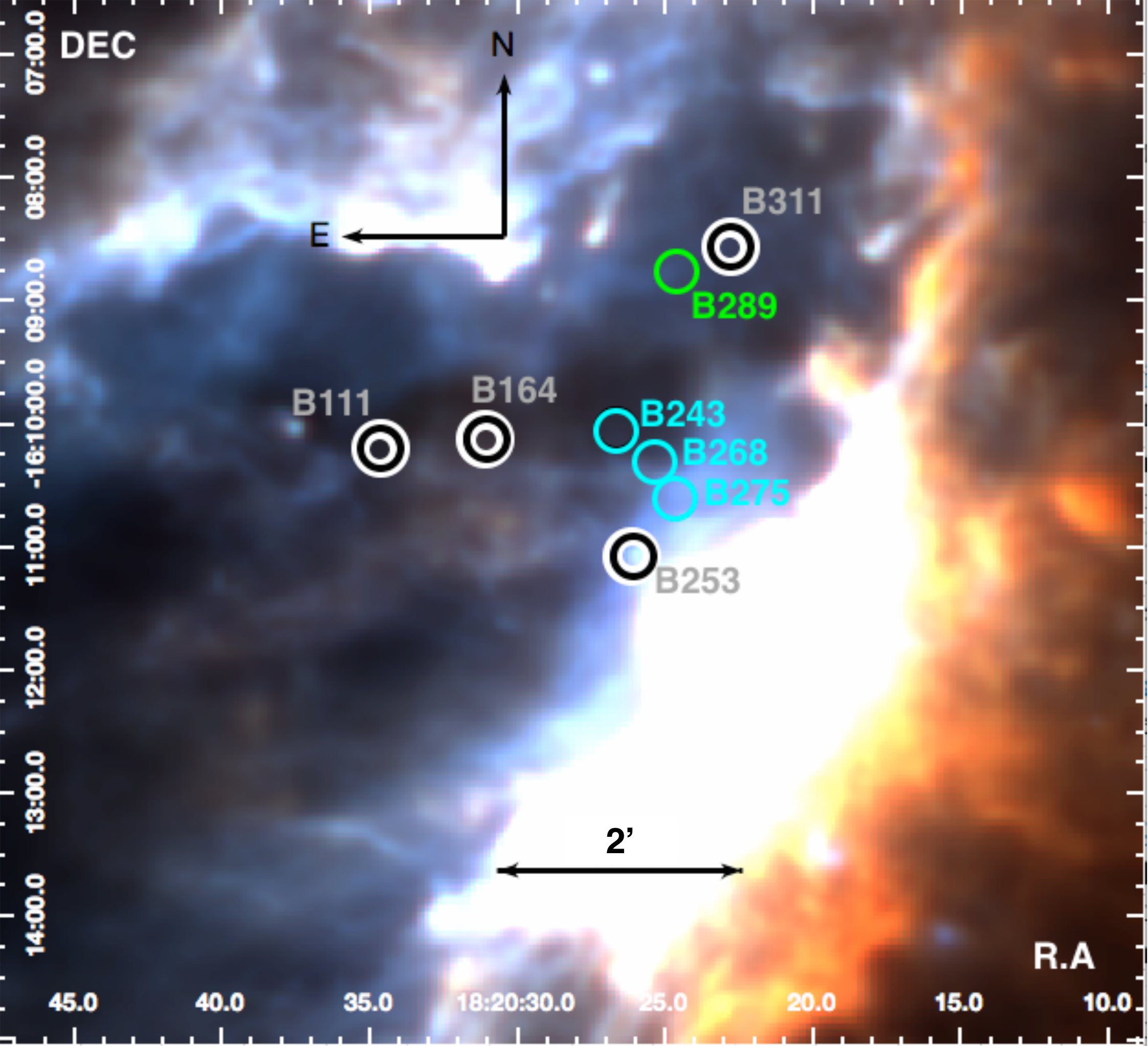}
      \caption{Far-infrared colour composite image of M17 based on {\it Herschel} PACS data. Blue: 70~$\mu$m, green: 100~$\mu$m, and red: 160~$\mu$m. The sources with circumstellar disks are shown in blue, the ones with IR excess longward of 2.5\,$\rm \mu m$ in green and the "naked" OB stars in black. We do not observe an obvious trend in extinction properties with the location in the region.}
         \label{P3:fig:Her_RGB}
   \end{figure}

We have obtained VLT/X-shooter ($300 - 2500$~nm) (\citealt{2011A26A...536A.105V}) spectra of 11 (pre-)main sequence (PMS) stars ranging in mass from 6 to 20~M$_\sun$. We selected this sample from the list of candidate massive PMS stars reported by \citet{1997ApJ...489..698H} and we determined their stellar and extinction properties, which are presented in \citetalias{paper1}. The location of the observed targets in M17 is shown in Fig.~\ref{P3:fig:Her_RGB}. With the exception of Fig.~\ref{P3:fig:EWAV_vs_Rv}, throughout this paper we color-code the M17 sightlines according to the properties of the stellar spectral energy distributions. The blue circles represent the sight lines towards objects with circumstellar disks (gas+dust disks), the green triangles towards stars with IR excess longwards of 2.5 $\mu$m (dusty disks), and the black squares towards OB stars without detected circumstellar material. We reduced the spectra using the X-shooter pipeline \citep{2010SPIE.7737E..28M} version 2.7.1 running under the ESO Reflex environment \citep{2013A&A...559A..96F} version 2.8.4. The flux calibration was obtained using spectrophotometric standards from the ESO database. We then scaled the NIR flux to match the absolutely calibrated VIS spectrum. The telluric correction was performed using the software tool \texttt{molecfit}  v1.2.0\footnote{\url{http://www.eso.org/sci/software/pipelines/skytools/molecfit}} \citep{2015A&A...576A..77S, 2015A&A...576A..78K}. The data reduction and analysis are described in detail in \citetalias{paper1}.

We calculated the stellar properties of the objects with visible photospheres by fitting FASTWIND models \citep{2005A&A...435..669P, 2012A&A...537A..79R} and obtained the total extinction in the $V$-band, $A_V$, and the visual-to-selective extinction ratio, $R_V$, by fitting the flux calibrated X-shooter spectra to \citet[][]{2004astro.ph..5087C} models.  The spectral type and extinction properties presented in \citetalias{paper1} are listed in Table~\ref{P3:tab:properties}.
In this paper we compare the results obtained in \citetalias{paper1} where we use the \citet{1989ApJ...345..245C} extinction law to deredden the flux calibrated spectra with results obtained by using the \citet{2007ApJ...663..320F} parameterisation of the extinction curve. We also compare the extinction parameters obtained by SED fitting with that derived from the colour excess E(B-V) (see Section~\ref{P3:sec:extinction}). We list the $R_V$ value obtained from Eq.~\ref{P3:eq:Rv_fitzpatrick}; for objects with a near-infrared excess (i.e., a circumstellar disk: B243, B268, and B275) we iterated this equation starting with $A_K=0$. In this way we obtain an estimate of the $K$-band extinction $A_K$, and thus a measure of the near-infrared excess (Section~\ref{P3:sec:extinction}). 

\begin{table*}
\centering
\caption{Extinction properties measured in the sight lines towards M17. The first column lists the star's identifier, the second the spectral type as reported in \citetalias{paper1}. The third, fourth, and fifth columns show the $B$, $V$, and $K$-band magnitudes, respectively. Columns 6, 8, and 10 list the extinction parameters ($R_V$, $A_V$, and $E(B-V)=A_V/R_V$) derived by fitting the spectral energy distribution to Kurucz models. We list the $R_V$ and $A_K$ values derived using Eq.~\ref{P3:eq:Rv_fitzpatrick} (see Section~\ref{P3:sec:extinction} for more information on objects with circumstellar disks) in columns 7 and 12. Column 9 lists the $V$-band excess $A_V$, and column 11 shows the colour excess ($B-V$)-($B-V$)$_0$, where the intrinsic colour ($B-V$)$_0$ is taken from \citet{2013ApJS..208....9P} and  \citet{2000MNRAS.319..771W, 2014AcA....64..261W}.
}
\begin{minipage}{\textwidth}
 \centering 
\renewcommand{\arraystretch}{1.4}
\setlength{\tabcolsep}{5pt}
\begin{tabular}{cccccccccccc}
\hline
\hline
Identifier & Sp.type    & $B$                                                                   & $V$                                                 & $K$\footnote{\label{P3:foot:2MASS}2MASS catalogue \citep{2003yCat.2246....0C}.} & $R_V$                     & $R_V$     & $A_V$                     & $A_V$     & $E(B-V)$          & $E(B-V)$                          & $A_K$ \\
           &            & mag                                                                   & mag                                                 & mag                                                                             & SED                       & Fitzp.    &  SED (mag)                & $V-V_0$   &   SED             & \scalebox{.6}{$(B-V)-(B-V)_0$}    &  mag  \\   
\hline
\hline
B111	&	O4.5 V		&	12.24\footref{P3:foot:AAVSO}	                                    &	11.21\footnote{\label{P3:foot:AAVSO}AAVSO Photometric all sky survey (APASS) catalog: \url{https://www.aavso.org/apass}}	& 7.475 & $3.7_{-0.3}^{+0.4}$ 	    & 3.86 	    &	$5.4_{-0.3}^{+0.4}$		& 5.17 &	$ 1.47 \pm 0.13 $	& 1.35  & 0.50 \\	
B164	&	O6 Vz		&	16.85\footnote{\label{P3:foot:Chini80}\citet{1980A&A....91..186C}} 	&	15.41\footref{P3:foot:Chini80}                                                                                              & 8.758 & $3.8_{-0.3}^{+0.3}$ 	    & 4.87 	    &	$8.3_{-0.5}^{+0.4}$ 	& 9.03 &	$ 2.2 \pm 0.10 $  	& 1.76  & 1.44 \\	
B215	&	B0-B1 V		&	$-$	                                                                &	16.10\footnote{\label{P3:foot:Hoff08}\citet{2008ApJ...686..310H}}                                                           & 10.00 & $4.1_{-0.2}^{+0.2}$ 	    & $-$  	    &	$7.6_{-0.3}^{+0.3}$ 	& 8.17 &	$ 1.85 \pm 0.06 $	& $-$   & 1.15 \\	
B243	&	B8 V		&	19.10\footref{P3:foot:Chini80}                                      &	17.80\footref{P3:foot:Chini80}                                                                                              & 9.544	& $4.7_{-0.8}^{\uparrow}$   & 4.64 	    &	$8.5_{-1.0}^{\uparrow}$	& 6.52 &	$ 1.82 \pm 0.22 $	& 1.41	& 0.72 \\	
B253	&	B3-B5 III	&	17.11\footref{P3:foot:Chini80} 	                                    &	15.74\footref{P3:foot:Chini80}                                                                                              & 10.31	& $4.1_{-0.2}^{+0.1}$ 	    & 4.29 	    &	$6.5_{-0.2}^{+0.1}$ 	& 6.22 &	$ 1.59 \pm 0.06 $	& 1.52  & 0.41 \\	
B268	&	B9-A0		&	18.40\footref{P3:foot:Chini80}  	                                &	17.10\footref{P3:foot:Chini80}                                                                                              & 9.494	& $4.6_{-0.8}^{\uparrow}$   & 3.47 	    &	$8.1_{-1.0}^{\uparrow}$	& 4.71 &	$ 1.75 \pm 0.25 $	& 1.34  & 0.52 \\	
B275	&	B7 III		&	17.02\footref{P3:foot:Chini80} 	                                    &	15.55\footref{P3:foot:Chini80}                                                                                              & 7.947	& $3.8_{-0.8}^{+0.7}$ 	    & 3.35 	    &	$6.7_{-1.0}^{+0.8}$ 	& 5.44 &	$ 1.77 \pm 0.26 $	& 1.59  & 0.61 \\	
B289	&	O9.7 V		&	16.97\footref{P3:foot:Chini80} 	                                    &	15.55\footref{P3:foot:Chini80}                                                                                              & 9.178	& $3.7_{-0.5}^{+0.4}$ 	    & 4.80 	    &	$8.3_{-0.8}^{+0.6}$ 	& 8.13 &	$ 2.23 \pm 0.17 $	& 1.73  & 0.78 \\	
B311	&	O8.5 Vz		&	14.99\footref{P3:foot:Chini80} 	                                    &	13.69\footref{P3:foot:Chini80}                                                                                              & 8.884 & $3.3_{-0.3}^{+0.3}$ 	    & 4.00 	    &	$6.1_{-0.4}^{+0.4}$ 	& 6.51 &	$ 1.85 \pm 0.11 $	& 1.62  & 0.71 \\	
B331	&	late-B		&	$-$ 	                                                            &	20.10\footref{P3:foot:Chini80}                                                                                              & 8.946 & $4.6_{-0.5}^{+0.5}$ 	    & $-$  	    &	$13.3_{-0.9}^{+0.9}$ 	& $-$  &    $ 2.89 \pm 0.13 $   & $-$   & $-$ \\	
B337	&	late-B		&	$-$ 	                                                            &	$-$                                                                                                                         & 9.343 & $3.7_{-0.5}^{+0.4}$ 	    & $-$  	    &	$8.3_{-0.8}^{+0.6}$ 	& $-$  &	$ 3.66 \pm 0.25 $	& $-$	& $-$ \\	
\hline
\end{tabular}

\renewcommand{\footnoterule}{}
\end{minipage}
\label{P3:tab:properties}
\end{table*}

\section{Extinction towards M17}
\label{P3:sec:extinction}

By dereddening the SEDs of a dozen OB stars \citet{1997ApJ...489..698H} detected a significant spread in both visual extinction $A_V$ (3--15 mag) and in total-to-selective extinction $R_{V}$ (2.8--5.5). They point out that the nebulosity disappears at the centre of the region and that the highest extinction is measured at the edges of this void. They also note that the DIB strengths (4430 and 4502~\AA) do not show a significant increase even though the visual extinction towards these objects varies from 3--10~mag. Earlier studies indicated that $R_V = 4.9$ \citep{1983A&A...117..289C} based on NIR photometry (but without knowledge of the spectral types), and $R_V = 3.3$ using the maximum polarisation relation for $R_V$ \citep{1981A&A....95...94S}.  \citet{2008ApJ...686..310H} concluded that the foreground extinction, with $A_{V} \sim 2$~mag and $R_{V} = 3.1$, differs from the extinction local to M17 for which they measure $R_{V} = 3.9 \pm 0.2$.

\subsection{Dust extinction}

In \citetalias{paper1} we constructed the spectral energy distribution (SED) of the stars by dereddening the X-shooter spectrum, as well as the available photometric data. In order to deredden the spectrum we used the \citet{1989ApJ...345..245C} extinction law. We then constrained $A_V$ and $R_{V}$ towards the objects in M17 by fitting the slope of their SED in the photospheric domain (400--820~nm) to Castelli \& Kurucz models\footnote{Table 2 in \url{http://www.stsci.edu/hst/observatory/crds/castelli_kurucz_atlas.html}} \citep{1993yCat.6039....0K, 2004astro.ph..5087C} corresponding to the spectral type reported in Table~\ref{P3:tab:properties}. 

The obtained values for $R_V$ range from 3.3 to 4.7 and $A_V$ varies from $\sim$6 to $\sim$15~mag. We do not find a significant correlation of the extinction properties with the position of the stars in the star forming region. \citet{1989ApJ...345..245C} mention that their parameterisation of the extinction law becomes inaccurate when $R_{V}$ is large. As an example they refer to the sight line to Herschel~36.

\begin{figure*}[t!]
\setlength{\tabcolsep}{2pt}
     \centering
        \subfigure{%
          \includegraphics[width=0.45\hsize]{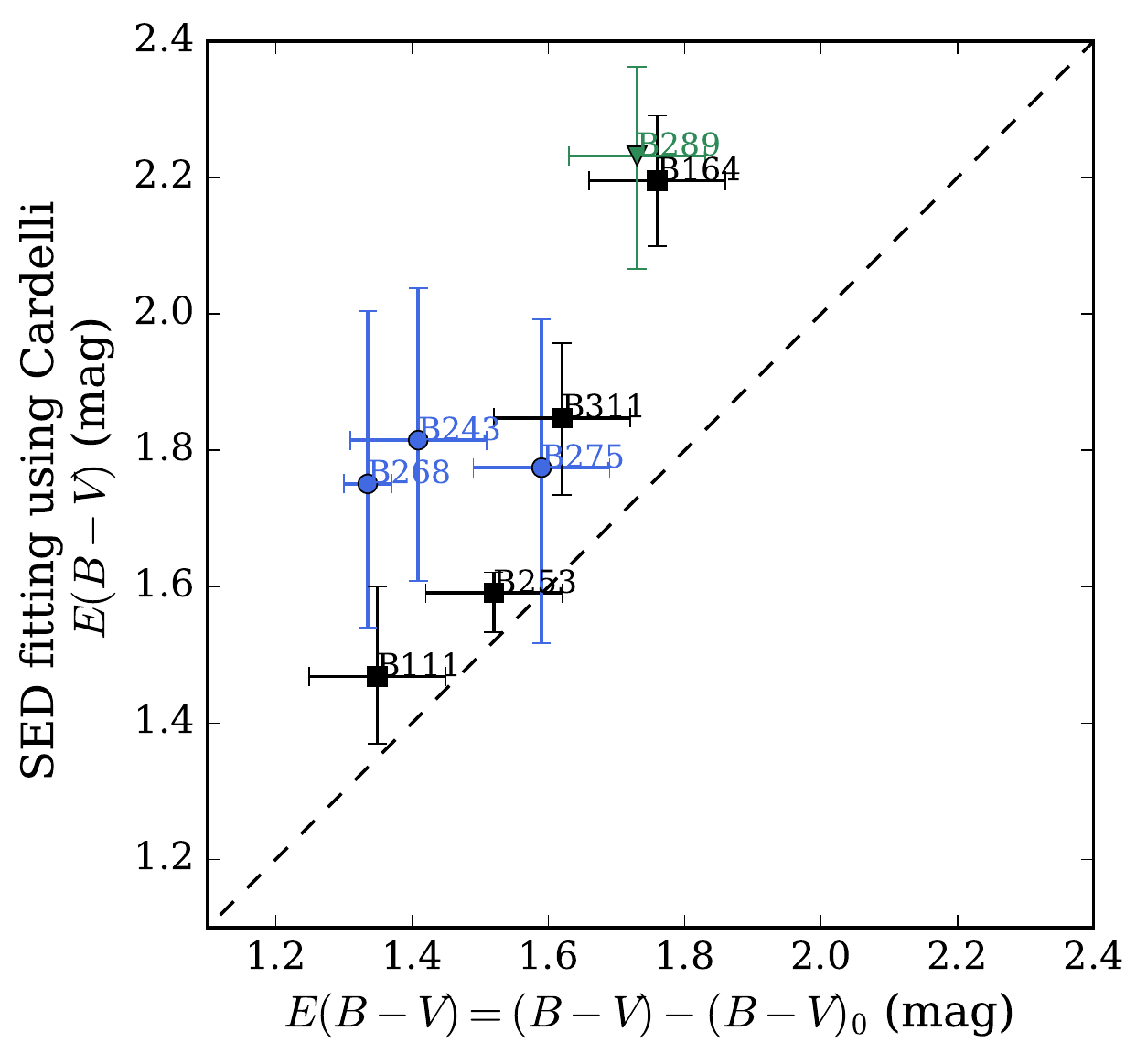}
        }%
        \hspace{5mm}
          \subfigure{%
            \includegraphics[width=0.45\hsize]{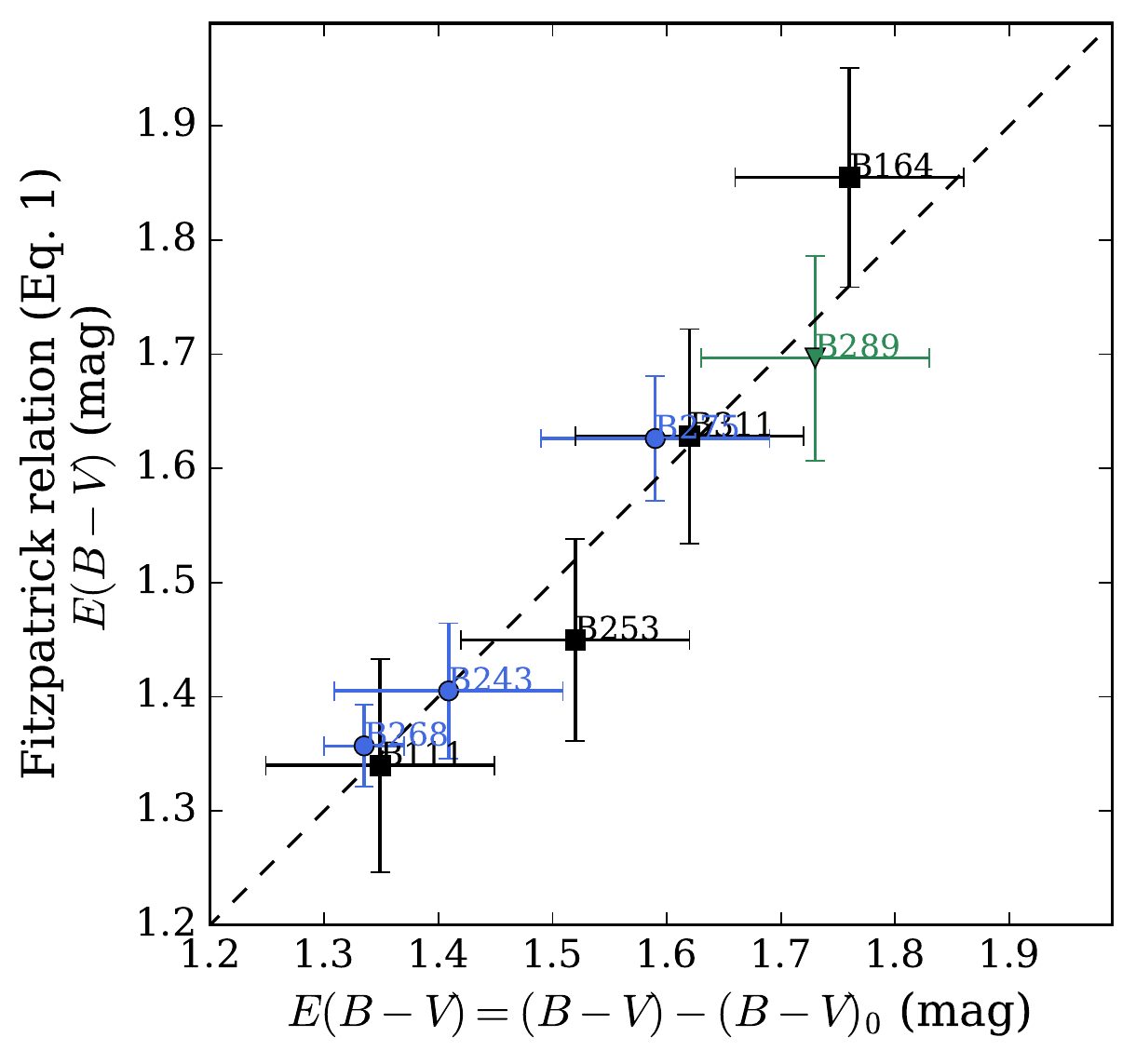}
        }\\%
    \caption[]{\emph{Left:} The colour excess $E(B-V)$ derived from SED fitting \citepalias[$A_V/R_V$, ][]{paper1} versus the value obtained using the observed and the intrinsic colour $(B-V)_0$; the latter is based on the spectral type reported in \citetalias{paper1}. \emph{Right:} The colour excess obtained from $A_V/R_V$, where $R_V$ is estimated using Eq.~1 in \citet{2007ApJ...663..320F}, versus the value obtained by using the observed and the intrinsic colour, as in the left panel. Now the discrepancy between these two values has been removed, demonstrating that the \citet{1989ApJ...345..245C} parameterisation underestimates the value of $R_{V}$ in these sight lines. The blue circles represent the sight lines towards  objects  with  circumstellar  disks, the green triangles towards stars with dusty disks, and the black squares towards OB stars.}
\label{P3:fig:E_BV_comparison}
\end{figure*}

As we know the spectral type of the targets, we can compare the value of the colour excess $E(B-V) \equiv A_V/R_V$ measured by fitting the SED with the value obtained directly by comparing the observed magnitudes to the intrinsic ones ($E(B-V)=(B-V)-(B-V)_0$). We adopt an error of 0.1~mag in $E(B-V)$ in order to account for the uncertainty in the spectral classification and the photometric errors. For the main sequence stars we took the value of $B_0$ and $V_0$ from \citet{2013ApJS..208....9P}. For the giants we used the calibration in \citet{2000MNRAS.319..771W, 2014AcA....64..261W}. The comparison is shown in the left panel of Fig.~\ref{P3:fig:E_BV_comparison}. The value of $E(B-V)$ derived from the SED fitting method is systematically higher than the one obtained directly from the observed and the intrinsic colours. The $E(B-V)$ values obtained with the SED fitting method and the \citet{1999PASP..111...63F} extinction law are closer to those obtained directly from the colors and we do not observe a systematic trend. Apparently, we underestimate the value of $R_{V}$ with the SED fitting method and the \citet{1989ApJ...345..245C} extinction law.

In order to investigate this further, we calculated $R_V$ using the relation reported in \citet{2007ApJ...663..320F}:

\begin{equation}
    R_V = -0.26 + 1.19\frac{E(V-K)}{E(B-V)}=-0.26 + 1.19\frac{A_V-A_K}{A_B-A_V}
\label{P3:eq:Rv_fitzpatrick}
\end{equation}

Given that this relation uses the $K-$band magnitude, it implicitly assumes that the target's SED does not include an IR excess or, alternatively, that this is corrected for. In the case of the three objects with circumstellar disks (blue dots in the plots), we calculated $R_V$ by iterating Eq.~\ref{P3:eq:Rv_fitzpatrick}. 
With $A_V=V-V_0$ and $E(B-V)=(B-V)-(B-V)_0$,
we started assigning a value of $A_K=0$ and calculated $R_V$ from this equation; we then used Fitzpatrick's extinction law to calculate $A_K$ with the obtained $R_V$ and repeated this process until the values for both $R_V$ and $A_K$ converged. 
We arrive at $R_V = 4.6, 3.3$ and $3.1$ for B243, B268, and B275, respectively\footnote{With this accurate measurement of the extinction properties we calculate $K$-band excesses of 2 to 3 magnitudes for B243, B268, and B275. In a forthcoming paper (Poorta et al., \emph{in prep.}) we investigate the physical properties of the circumstellar disks that cause the measured $K$-band excess.}. The comparison between the colour excess calculated with Fitzpatrick's extinction law and the one based on the observed and intrinsic colour is shown in the right panel of Fig.~\ref{P3:fig:E_BV_comparison}. From the figure it is clear that the values obtained using these two methods agree within the errors. 
We note that, in the case of objects with near-IR excess, these two methods are not completely independent, because we used the values based on the observed and intrinsic colour as initial guess to iterate Eq.~\ref{P3:eq:Rv_fitzpatrick}.

We conclude that with the method used in \citetalias{paper1}, where the \citet{1989ApJ...345..245C} extinction law is used to deredden the SEDs for sightlines with high $R_V$, we underestimate the value of this parameter. Furthermore, we observe that the values for $A_V$ obtained with the three methods agree with each other. This means that we do not make a significant error when calculating the luminosity of the objects in \citetalias{paper1}.

We do not find a systematic difference between the values obtained for objects with circumstellar gaseous disks (blue dots), only dusty disks (green triangle), and naked OB stars (black squares). 

In the remainder of this paper we will adopt the values of $A_V = V-V_0$, $E(B-V)=(B-V)-(B-V)_0$ and $R_V$ derived from Eq.~\ref{P3:eq:Rv_fitzpatrick}.

\subsection{The gaseous component of the extinction}

\begin{figure}[t!]
    \centering
    \includegraphics[width=\hsize]{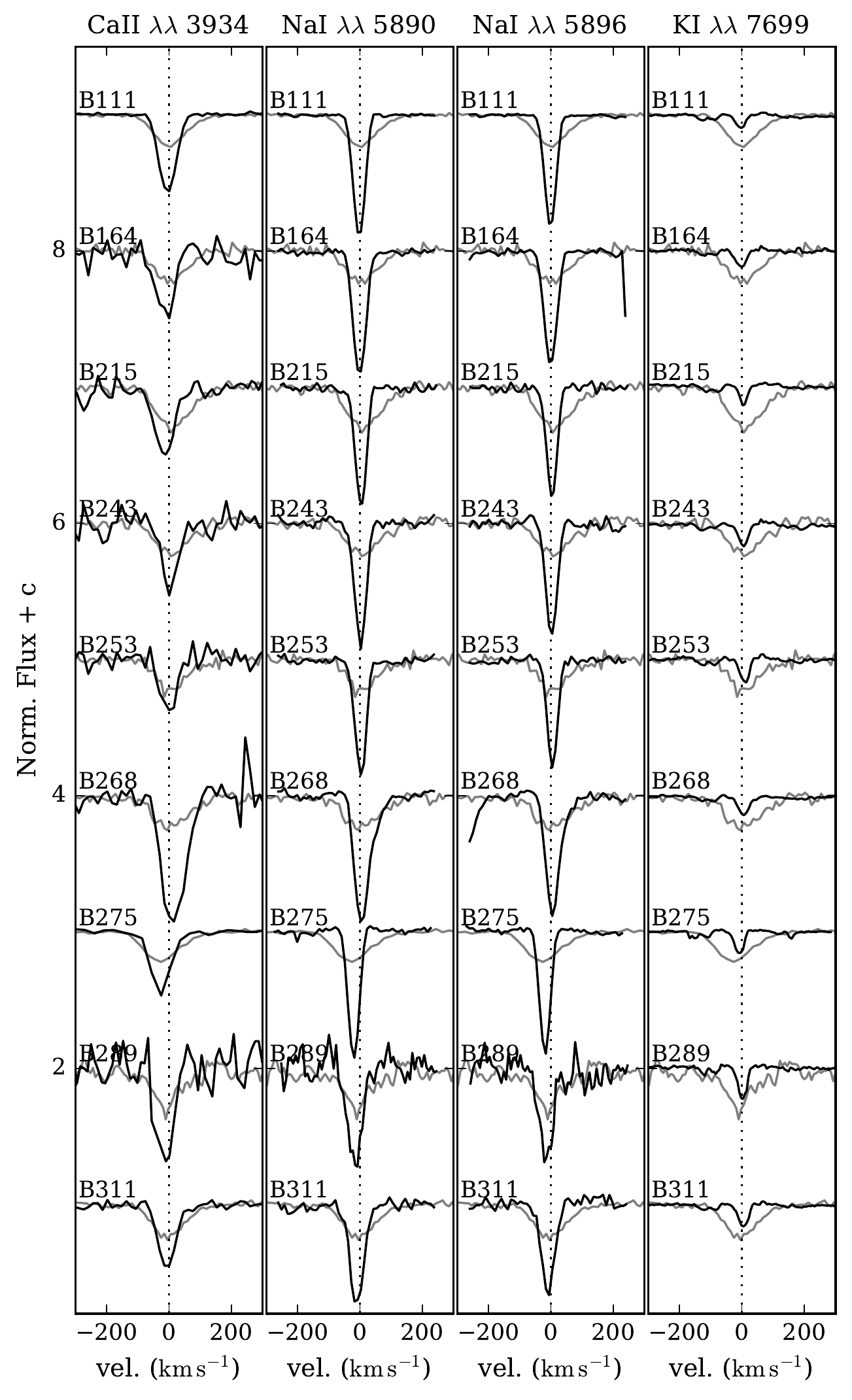}
    \caption{Velocity profiles for the \ion{Ca}{ii}, \ion{Na}{I}, and \ion{K}{i} lines (black). We have over plotted the DIB at 5780~\AA\ for reference (grey). The velocities are in the barycentric rest frame. The interstellar (gaseous) absorption is very similar comparing one sight line to the other and is likely due to a single cloud.
    }
    \label{P3:fig:NaICaIIKI}
\end{figure}

The \ion{Ca}{ii}, \ion{Na}{i}, and \ion{K}{i} interstellar lines are shown in Figure~\ref{P3:fig:NaICaIIKI}; we have over plotted the DIB at 5780~\AA. The velocities have been corrected into the barycentric rest frame. In Table~\ref{P3:tab:rvNaICaIIKI_LSR} we list the local standard rest frame (LSR) velocity measured for each of the lines. These radial velocities are consistent with interstellar absorption at distances of 1.3--2.2~kpc based on Galactic rotation models by \citet{2009ApJ...700..137R}. We observe hardly any difference from one sight line to the other. The atomic lines 
are mostly unresolved at the moderate X-Shooter spectral resolution of $R \sim 11000$. Although inconclusive regarding the exact distribution of gas in these lines-of-sight these data indicate the main gaseous absorption features are associated with M17 at $\sim$2~kpc.

From 3D dust extinction maps \citep{2003A&A...409..205D} we find that the predicted visual extinction up to 1.8~kpc in the direction of M17 (i.e. in the foreground) corresponds to roughly 2~mag (or $E(B-V) = 0.65$~mag for a canonical value $R_V = 3.1$), which is consistent with \citet{2008ApJ...686..310H}. From the $E(B-V)$ versus distance towards M17 plot obtained via the online tool \textsc{Stilism}\footnote{\url{http://stilism.obspm.fr}} by \citet{2017A&A...606A..65C} we obtain a value of $E(B-V)=0.78$~mag at the distance M17, which is in line with our result. This plot shows that there are diffuse clouds in the foreground to the local gas and dust associated with the M17 region.

\begin{table}
\setlength{\tabcolsep}{3pt}
\caption{Local standard rest frame (LSR) radial velocities obtained from a Gaussian fit to the \ion{Ca}{ii}, \ion{Na}{i}, and \ion{K}{i} interstellar lines in each line of sight.}
\label{P3:tab:rvNaICaIIKI_LSR}
\centering
\begin{tabular}{ccccc}\hline\hline
Identifier    & \ion{Ca}{ii} &  \ion{Na}{i} & \ion{Na}{i}  & \ion{K}{i} \\
$\lambda\lambda$ & 3933.66 & 5889.95 & 5895.92 & 7698.974 \\
    &   $\rm km\,s^{-1}$ &   $\rm km\,s^{-1}$ &   $\rm km\,s^{-1}$    & $\rm km\,s^{-1}$ \\
\hline
B111 & $  9.3 \pm0.4$ & $ 11.3\pm0.2$ & $12.1\pm0.2$ & $-$         \\
B164 & $  0.8 \pm3.5$ & $ 12.6\pm0.3$ & $12.7\pm1.3$ & $10.4\pm1.3 $ \\
B215 & $ -1.2\pm2.3$ & $ 16.8\pm0.3$ & $16.7\pm0.3$ & $18.9\pm1.0 $ \\
B243 & $ 16.6\pm13.5$ & $ 15.4\pm0.4$ & $15.7\pm0.5$ & $18.2\pm1.4 $ \\
B253 & $ 10.7\pm6.1$ & $ 16.2\pm0.4$ & $18.6\pm0.3$ & $22.4\pm1.1 $ \\
B268 & $ 28.8\pm2.2$ & $ 21.7\pm0.8$ & $20.4\pm1.4$ & $19.6\pm1.5 $ \\
B275 & $-13.2\pm1.0$ & $ -6.1\pm0.4$ & $-5.1\pm0.3$ & $ 5.1\pm1.1 $ \\
B289 & $ -9.1\pm5.2$ & $ -11.3\pm1.6$ & $-9.9\pm1.7$ & $-$ \\
B311 & $  7.3 \pm1.1$ & $  4.1\pm0.7$ & $ 4.6\pm0.8$ & $-$ \\
\hline
\end{tabular}
\end{table}

\section{Diffuse interstellar bands}
\label{P3:sec:DIBs}

In this section we provide an overview of the DIB properties measured in the sight lines to the young OB stars in the star forming region M17.  
We analysed the nine strongest DIBs present in the spectra obtained by the UVB and VIS arms of X-shooter (at 4430, 5780, 5797, 6196, 6284, 6379, 6614, 7224, and 8620~\AA). We also included the NIR DIBs at 11797 and 13176, first reported by \citet{1990Natur.346..729J}, and the one at 15268 (\citealt{2011Natur.479..200G}; \citealt{2014A&A...569A.117C}). 
The DIB equivalent width (EW) and full width at half maximum (FWHM) are listed in Table \ref{P3:tab:DIBs}. The first column corresponds to the central wavelength of the DIBs, the top values show the FWHM in nm, and the bottom values show the EW measurements in (m\AA). 

We also detect the 9577 and 9632 DIBs whose carrier has recently been identified in the laboratory to be the $\rm C^{+}_{60}$ molecule \citep[][]{2015Natur.523..322C}. The FWHM and EW of these two DIBs are displayed in rows 10 and 11 of Table~\ref{P3:tab:DIBs}.

We searched for the presence of CH$^+$ and CH at 4232, 4300~\AA\, but given the signal to noise of our data at such blue wavelengths, we were not able to detect these molecules in the X-shooter spectra.

\subsection{DIB strengths and widths}

\begin{table*}
\setlength{\tabcolsep}{3pt}
\renewcommand{\arraystretch}{1.4}
\caption{DIB properties in the lines-of-sight to M17. The first column shows the rest wavelength of the DIBs. Columns 2 to 12 show the FWHM~(nm; first row) and EW (m\AA; second row) measured for the sight lines towards the objects in M17.}
\label{P3:tab:DIBs}
\scalebox{0.85}{\begin{tabular}{llllllllllll}
\toprule
              $\lambda$ &           B111 &            B164 &            B215 &            B243 &            B253 &            B268 &           B275 &            B289 &           B311 &          B331 &           B337 \\
\midrule
  \multirow{2}{*}{4430} &         $1.94$ &           $2.1$ &          $2.93$ &          $1.69$ &          $2.02$ &          $1.89$ &          $1.8$ &          $2.33$ &         $2.31$ &            -- &             -- \\
                        &  $2027 \pm 32$ &  $2539 \pm 145$ &  $4794 \pm 218$ &  $1779 \pm 240$ &  $2020 \pm 158$ &  $1742 \pm 130$ &  $1919 \pm 36$ &  $3304 \pm 311$ &  $2528 \pm 53$ &            -- &             -- \\
  \midrule
\multirow{2}{*}{5780} &         $0.23$ &          $0.22$ &          $0.24$ &          $0.23$ &          $0.21$ &          $0.27$ &         $0.25$ &          $0.24$ &         $0.23$ &            -- &             -- \\
                        &   $558 \pm 12$ &    $521 \pm 38$ &    $772 \pm 34$ &    $540 \pm 45$ &    $507 \pm 38$ &    $663 \pm 49$ &   $558 \pm 17$ &    $711 \pm 49$ &   $596 \pm 24$ &            -- &             -- \\
\midrule
  \multirow{2}{*}{5797} &         $0.11$ &          $0.14$ &              -- &              -- &              -- &          $0.26$ &         $0.17$ &              -- &         $0.15$ &            -- &             -- \\
                        &     $89 \pm 8$ &    $129 \pm 32$ &              -- &              -- &              -- &    $185 \pm 30$ &    $125 \pm 8$ &              -- &    $168 \pm 8$ &            -- &             -- \\
\midrule
  \multirow{2}{*}{6196} &         $0.09$ &          $0.08$ &          $0.07$ &          $0.08$ &          $0.09$ &          $0.06$ &          $0.1$ &              -- &         $0.09$ &            -- &             -- \\
                        &     $60 \pm 6$ &     $62 \pm 12$ &     $60 \pm 11$ &     $58 \pm 13$ &     $51 \pm 13$ &     $55 \pm 13$ &     $64 \pm 8$ &              -- &     $63 \pm 6$ &            -- &             -- \\
\midrule
  \multirow{2}{*}{6284} &         $0.35$ &          $0.29$ &          $0.42$ &          $0.25$ &          $0.26$ &          $0.33$ &         $0.27$ &          $0.35$ &         $0.28$ &            -- &             -- \\
                        &  $1691 \pm 11$ &   $1790 \pm 24$ &   $2206 \pm 28$ &   $1720 \pm 34$ &   $1708 \pm 38$ &   $1580 \pm 29$ &  $1661 \pm 36$ &   $1920 \pm 57$ &  $1770 \pm 25$ &            -- &             -- \\
\midrule
  \multirow{2}{*}{6379} &          $0.1$ &          $0.09$ &          $0.13$ &           $0.2$ &          $0.08$ &          $0.07$ &         $0.08$ &          $0.08$ &         $0.09$ &            -- &             -- \\
                        &    $35 \pm 12$ &     $51 \pm 10$ &     $80 \pm 15$ &     $54 \pm 24$ &     $35 \pm 13$ &      $32 \pm 8$ &     $47 \pm 5$ &     $65 \pm 10$ &    $64 \pm 12$ &            -- &             -- \\
\midrule
  \multirow{2}{*}{6614} &         $0.14$ &          $0.14$ &          $0.13$ &          $0.13$ &          $0.13$ &          $0.14$ &         $0.14$ &          $0.14$ &         $0.13$ &            -- &             -- \\
                        &    $188 \pm 6$ &    $203 \pm 27$ &    $251 \pm 12$ &    $187 \pm 19$ &    $168 \pm 15$ &    $201 \pm 20$ &   $203 \pm 27$ &    $243 \pm 29$ &   $227 \pm 11$ &            -- &             -- \\
\midrule
  \multirow{2}{*}{7224} &         $0.16$ &          $0.15$ &          $0.18$ &          $0.17$ &          $0.15$ &          $0.17$ &         $0.18$ &          $0.15$ &         $0.17$ &        $0.16$ &         $0.14$ \\
                        &   $313 \pm 10$ &    $287 \pm 12$ &    $415 \pm 11$ &     $386 \pm 9$ &    $305 \pm 16$ &    $339 \pm 12$ &   $298 \pm 10$ &    $337 \pm 15$ &   $376 \pm 11$ &  $387 \pm 28$ &  $373 \pm 233$ \\
\midrule
  \multirow{2}{*}{8620} &             -- &          $0.53$ &          $0.48$ &          $0.47$ &          $0.42$ &          $0.49$ &         $0.44$ &          $0.43$ &         $0.55$ &        $0.48$ &         $0.54$ \\
                        &             -- &    $562 \pm 17$ &    $471 \pm 14$ &    $485 \pm 14$ &    $370 \pm 12$ &    $493 \pm 12$ &   $400 \pm 12$ &    $457 \pm 19$ &    $506 \pm 5$ &  $494 \pm 16$ &   $630 \pm 20$ \\
\midrule
  \multirow{2}{*}{9577} &         $0.44$ &          $0.45$ &          $0.48$ &          $0.47$ &          $0.55$ &          $0.45$ &         $0.42$ &          $0.45$ &         $0.45$ &         $0.4$ &         $0.47$ \\
                        &   $371 \pm 14$ &    $579 \pm 11$ &    $546 \pm 22$ &    $479 \pm 13$ &    $512 \pm 20$ &    $435 \pm 11$ &   $439 \pm 10$ &    $526 \pm 13$ &   $632 \pm 17$ &  $328 \pm 23$ &   $489 \pm 16$ \\
\midrule
  \multirow{2}{*}{9632} &         $0.38$ &          $0.39$ &          $0.46$ &          $0.42$ &          $0.39$ &          $0.46$ &          $0.4$ &          $0.38$ &         $0.36$ &        $0.29$ &         $0.52$ \\
                        &   $401 \pm 33$ &    $586 \pm 26$ &    $636 \pm 42$ &    $499 \pm 34$ &    $509 \pm 47$ &    $531 \pm 25$ &   $513 \pm 27$ &    $550 \pm 24$ &   $649 \pm 27$ &  $436 \pm 31$ &   $613 \pm 39$ \\
\midrule
 \multirow{2}{*}{11797} &         $0.32$ &          $0.27$ &          $0.32$ &          $0.39$ &          $0.31$ &          $0.36$ &         $0.21$ &          $0.27$ &         $0.26$ &        $0.64$ &         $0.99$ \\
                        &   $278 \pm 13$ &    $262 \pm 14$ &    $205 \pm 15$ &    $306 \pm 33$ &    $269 \pm 29$ &    $336 \pm 18$ &   $173 \pm 44$ &    $255 \pm 20$ &   $261 \pm 14$ &  $318 \pm 31$ &   $457 \pm 49$ \\
\midrule
 \multirow{2}{*}{13176} &         $0.55$ &          $0.62$ &          $0.67$ &          $0.67$ &          $0.62$ &          $0.64$ &         $0.35$ &          $0.54$ &         $0.53$ &        $0.53$ &         $0.73$ \\
                        &   $923 \pm 82$ &   $1134 \pm 47$ &   $1161 \pm 56$ &  $1115 \pm 108$ &   $1044 \pm 84$ &   $1016 \pm 76$ &   $463 \pm 70$ &    $928 \pm 71$ &   $859 \pm 53$ &  $796 \pm 60$ &  $1635 \pm 88$ \\
\midrule
 \multirow{2}{*}{15268} &         $0.46$ &          $0.71$ &          $0.58$ &              -- &          $0.74$ &          $0.57$ &             -- &          $0.52$ &         $0.51$ &            -- &             -- \\
                        &   $395 \pm 34$ &    $875 \pm 25$ &    $521 \pm 42$ &              -- &    $615 \pm 47$ &    $820 \pm 42$ &             -- &    $438 \pm 18$ &   $401 \pm 17$ &            -- &             -- \\
\bottomrule
\end{tabular}
}
\end{table*}

The EW was measured by integrating the flux of each DIB within the boundaries shown in Appendix~\ref{P3:appendix_DIBs}. 
The error in the EW was calculated following the method described in \citet{2011A&A...533A.129V}:
$\sigma_\mathrm{EW}=\sqrt{2 \Delta\lambda \delta\lambda} / \mathrm{SNR}$, where $\Delta\lambda$ is the integration range and $\delta\lambda$ is the spectral dispersion. The real errors might be larger due to unknown systematic uncertainties, particularly regarding the continuum normalisation. The SNR was measured in the shaded region shown in Appendix~\ref{P3:appendix_DIBs}. The FWHM ($=2\sqrt{2 \ln{2}}\, \sigma$) of each DIB was measured by fitting a Gaussian function to the absorption lines.

\subsection{Correlation between EW(6196) and EW(6613)}

\begin{figure}
    \centering
    \includegraphics[width=0.95\hsize]{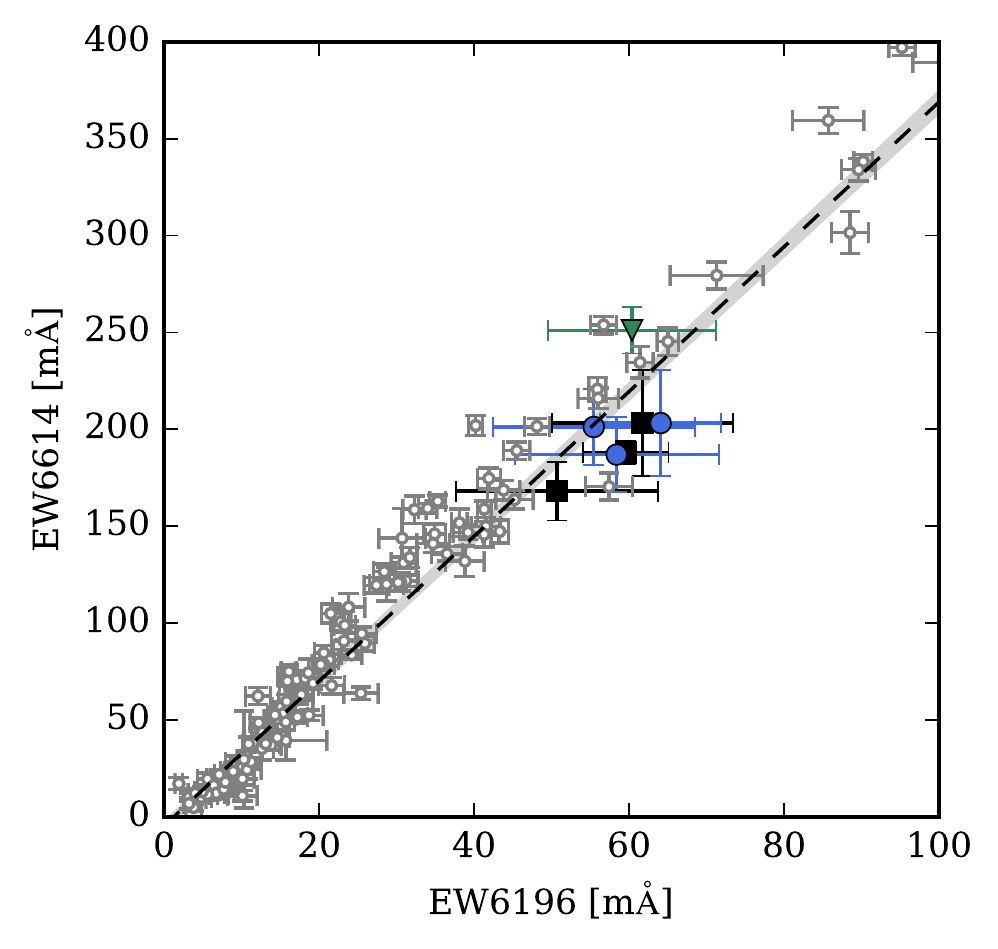}
    \caption{Correlation between the DIBs at 6614 and 6196 \AA. The dashed line represents the relation found by \citet{2010ApJ...708.1628M} (grey dots) and the grey area shows the 1$\sigma$ errors to the fit. 
    /textbf{The blue circles represent the sight lines towards objects with circumstellar disks, the green triangles towards stars with IR excess longwards of 2.5 $\mu$m, and the black squares towards OB stars.}
    }
    \label{P3:fig:6614_vs_6696}
\end{figure}

In order to check the consistency of the measurements we show our data, together with literature data, for the strongly correlated DIBs at 6614 and 6696~\AA\  in Fig.~\ref{P3:fig:6614_vs_6696} \citep[][]{2010ApJ...708.1628M, 2016MNRAS.460.2706K}. The dashed line shows a linear fit to the data by \citet{2010ApJ...708.1628M} and the grey region shows the 2$\sigma$ error on the fit. The measured ratios are in agreement with the trend observed in other Galactic sight lines.

\subsection{C$_{60}^+$ bands}

We measured the C$_{60}^+$ bands at 9577 and 9632~\AA. The EW of the 9577 DIB increases roughly with $A_V$.  B331 is an outlier with relatively weak 9577 and 9632~\AA\ DIBs. The average measured 9632/9577 ratio is 1.1 which is higher than the ratio 0.8 expected from laboratory measurements. The 9632~\AA\ DIB is known to suffer from contamination by a photospheric \ion{Mg}{ii} line.
However, unless there are other factors impacting this ratio in M17, the stellar contamination has to be, on average, higher than 150~m\AA\ in order to reduce the observed ratio to 0.8. This appears to require a rather large contribution from photospheric \ion{Mg}{ii} and we note that we cannot exclude systematic errors originating from telluric residuals. 

In the optically thin limit the EW of the 9577~\AA\ DIB can be converted into a column density. We do not use the 9632~\AA\ DIB as it is known to suffer from photospheric line contamination which is difficult to correct for.
For the observed 9577~\AA\ DIB strengths this gives N(C$_{60}^+$) = 2.5--4.3 $\times 10^{13}$~cm$^{-2}$. 
These are somewhat lower abundances than the neutral C$_{60}$ abundance of 2.4 $\times$ 10$^{14}$~cm$^{-2}/G_0$, where $G_0$ is the radiation field, measured with \textit{Spitzer} in the diffuse ISM \citep{2017A&A...605L...1B}. However, the higher sensitivity of JWST opens the possibility to detect fullerene infrared emission (potentially both C$_{60}$ and C$_{60}^+$ features) towards M17 that would allow a comparison between emission and absorption measurements in the same environment.

\subsection{DIB profiles}

\citet{2013ApJ...773...41D} reported on anomalous DIB profiles in the sight line towards Herschel~36. According to the authors, the profiles of the 5780, 5797, 6196, and 6614 DIBs show an extended red wing in comparison to nominal sight lines. \citet{2013ApJ...773...42O} confirmed this finding for three of the DIBs, but did not detect a pronounced red wing in the DIB at 6196~\AA\ (we also do not detect a red wing in the profile of the 6196 DIB with the lower-resolution X-shooter data). These authors performed a study that allowed them to make a distinction between the carriers of the above DIBs at 5780, 5797, and 6614 DIBs and others (e.g. 5850, 6196, and 6379) which do not present such a pronounced red wing. 

In addition, the sight line towards Herschel~36 shows absorption lines from rotationally excited CH$^{+}$ and CH and from vibrationally excited H$_{2}$. Furthermore, an atypical extinction curve is observed with $R_V\sim6$ and a weak 2175~\AA\ bump, i.e., a flat far-UV extinction curve  \citep[][]{2013ApJ...773...41D}. M17 also shows anomalous extinction \citepalias[$R_V>3.1$,][]{paper1}, therefore, it is a good candidate to search for anomalous behaviour in the DIBs.

\begin{figure}
    \centering
    \includegraphics[width=0.98\hsize]{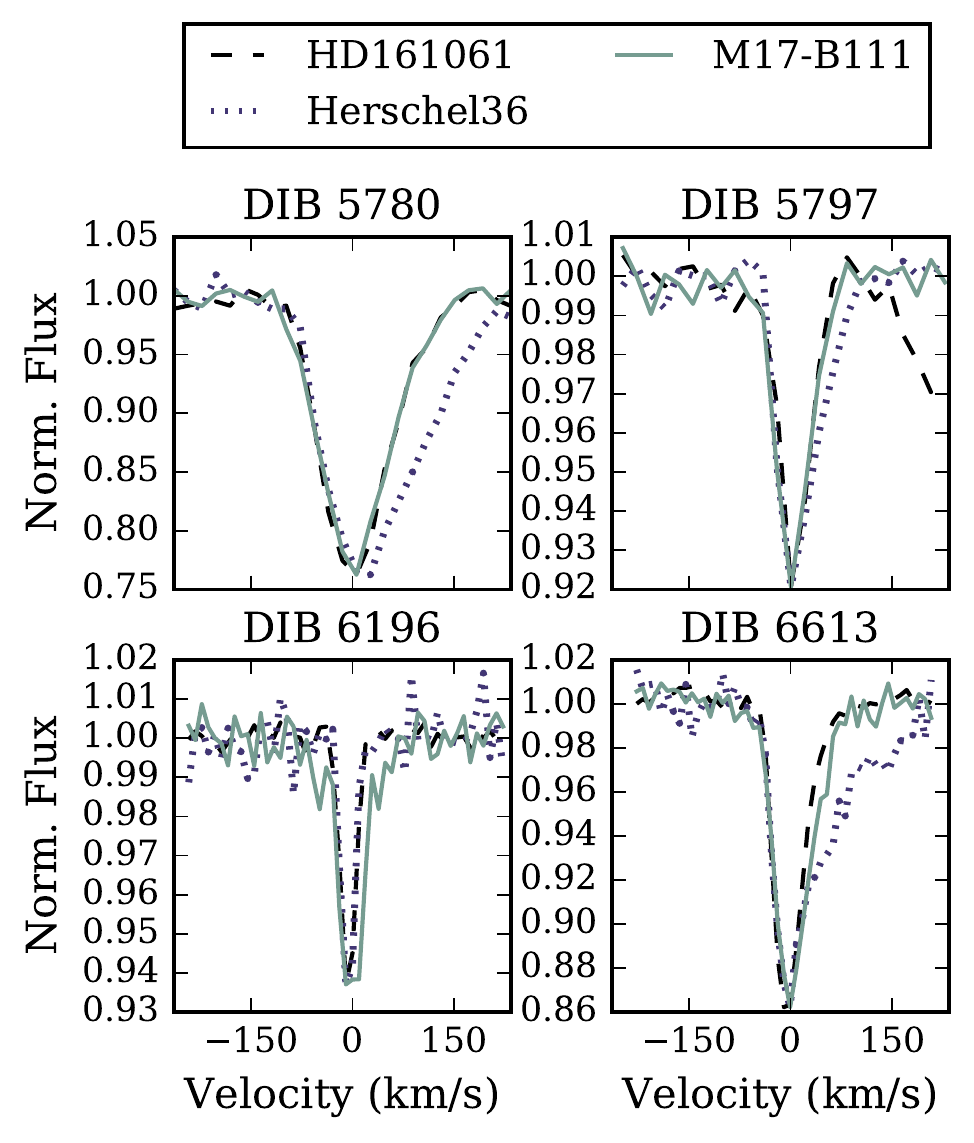}
    \caption{Comparison of the shape of some DIBs towards M17 (B111, green, solid) with the sight line towards Herschel~36 (purple, dotted) and a 'nominal' sight line (black, dashed). All the spectra have been convolved to the same resolution and scaled to have the same line central depth. The M17 DIB profiles do not show an extended red wing, nor another anomaly with respect to the DIB in HD161061.
    }
    \label{P3:fig:compfourDIBs}
\end{figure}

We compare the profiles of the DIBs towards M17 with the ones observed in the direction of Herschel~36 and HD~161061 (a nominal sight line) in Fig.~\ref{P3:fig:compfourDIBs}. The X-shooter spectra of Herschel~36 (unpublished) and the nominal sight line (HD~161061) were obtained previously by us with ESO programs 091.C-0934(B) and 385.C-0720(A). The spectra for these two sources were reduced using the same procedure described in Sect.~\ref{P3:sec:obs}. 

To facilitate the comparison, we have scaled the flux of the DIBs towards Herschel~36 and HD~161061 to match the depths of the cores of the DIBs towards M17. The profiles of the DIBs towards M17 are similar to the 'nominal' profile, and they do not show a red wing like in the case of Herschel~36. In the case of the 6196~\AA\  DIB we do not detect a red wing in the spectrum of Herschel~36.

\section{Comparison with other sight lines}
\label{P3:sec:OtherSightlines}

For reddened lines of sight the strength of many DIBs increases in a roughly linear way with E($B-V$). 
The relation between the DIB strength and the reddening is not very well constrained for large extinction values because of the lack of data covering those high values of E($B-V$). Also, these studies have been applied to only a few prominent DIBs, mostly the 5797, 5780 and 6614~\AA\  DIBs. 

\begin{figure*}[ht!]
    \centering
    \subfigure{
        \includegraphics[width=0.45\hsize]{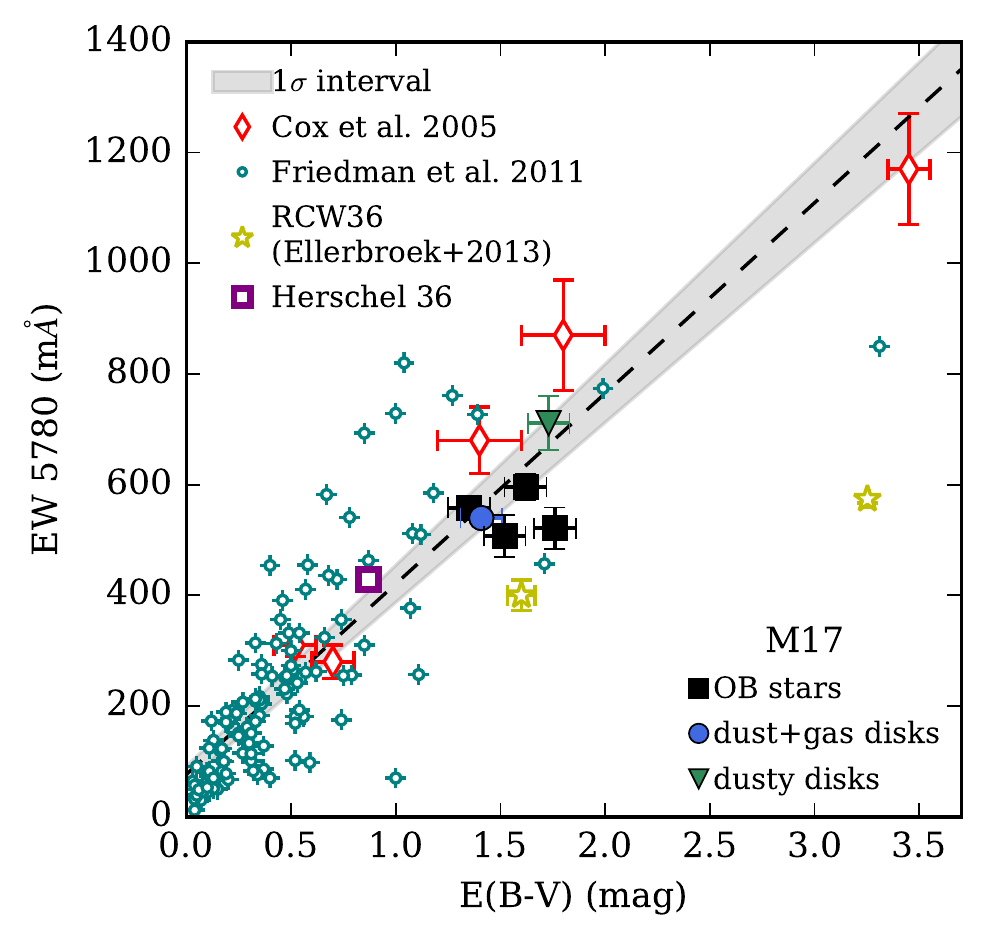}
    }
    \subfigure{
        \includegraphics[width=0.45\hsize]{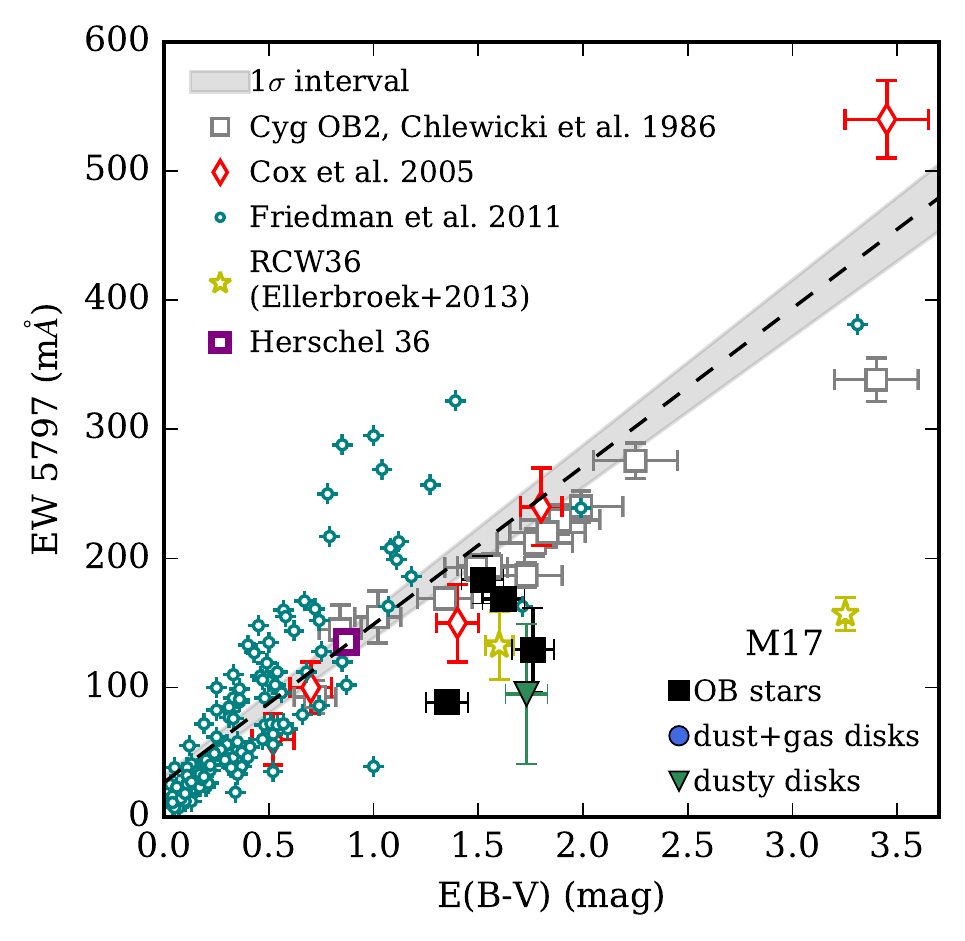}
    }\\
    \subfigure{
        \includegraphics[width=0.45\hsize]{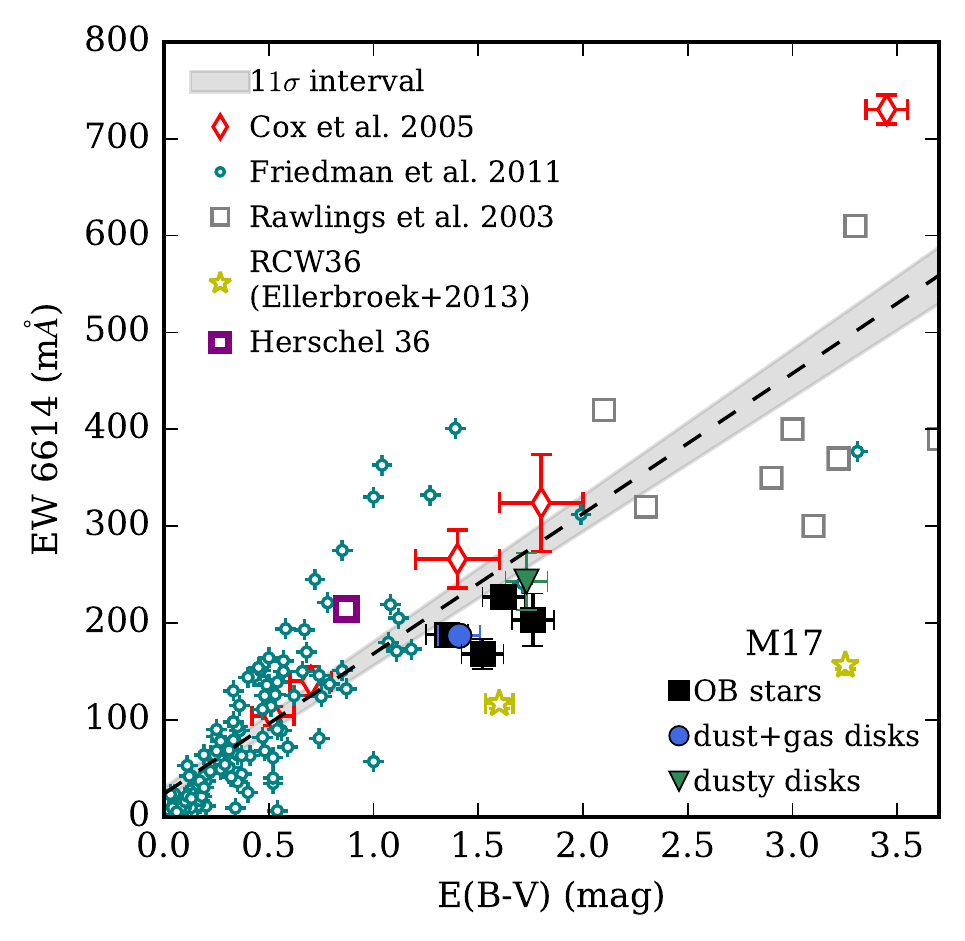}
    }
    \caption{Equivalent width of the 5780 (top left), 5797 (top right), and 6614~\AA\  (bottom) DIBs as a function of reddening E($B-V$) for Galactic sightlines. The DIB strengths are taken from the literature (open symbols) and this work (filled symbols). The dashed line indicates the linear fit for the Galactic sight lines excluding the star forming regions (RCW\,36, Herschel 36, and M17) and the grey area represents the $1\sigma$ error on the fit.
    }
    \label{P3:fig:DIBvsEBV}
\end{figure*}

In Fig.~\ref{P3:fig:DIBvsEBV} we compare the behaviour of some prominent DIBs in M17 as a function of the reddening \mbox{E($B-V$)} with that in other published studies. 
We plot a linear fit to a large variety of Galactic sightlines \citep[][for details see figure captions]{1986ApJ...305..455C,1989A&A...223..329B,1993ApJ...407..142H,1997A&A...327.1215S,1999A&A...347..235K,2000A&AS..142..225T,2003MNRAS.341.1121R,2004MNRAS.355..169G,2005A&A...438..187C,2011ApJ...727...33F} excluding the star forming regions (M17, Herschel~36, and RCW~36) for the three DIBs mentioned above with dashed black lines.
With the grey area we show the linear fit and the 1$\sigma$ error bars to the fit. For reference, we also plot DIB equivalent width measurements for Herschel~36 and for the massive star-forming region RCW 36 \citep{2013A&A...551A...5E}.
The large scatter on these relations is intrinsic due to physical variations between lines-of-sight, and not due to measurement uncertainties.
For example, \citealt{2011A&A...533A.129V} show that there is a dichotomy in the DIB strength versus extinction relation for UV exposed and UV shielded environments.
 
The M17 points are located below the average Galactic trend, like the sightlines probing the dense Cyg~OB2 region. 
On the other hand, when including the DIBs towards M17 in the fit, the slope of the relation does not change significantly for each of the three DIBs.

We also compared the strength of the NIR DIB at 15268~\AA\  as a function of E($B-V$) with several Galactic sight lines taken from the SDSS-III APOGEE survey by \citet{2014IAUS..297...68Z}. 
We do not show this comparison, but we observe that the sight lines towards M17 follow the same trend as the ones observed in different Galactic environments. 
We conclude that the DIB behaviour with respect to reddening \mbox{E($B-V$)}, as observed in the direction of M17, indicates that these sightlines probe a denser ISM, but are not peculiar.

\subsection{Ratio EW(5797)/EW(5780)}
\label{P3:sec:ratio_5797_5780}

\begin{figure}
    \centering
    \includegraphics[width=0.95\hsize]{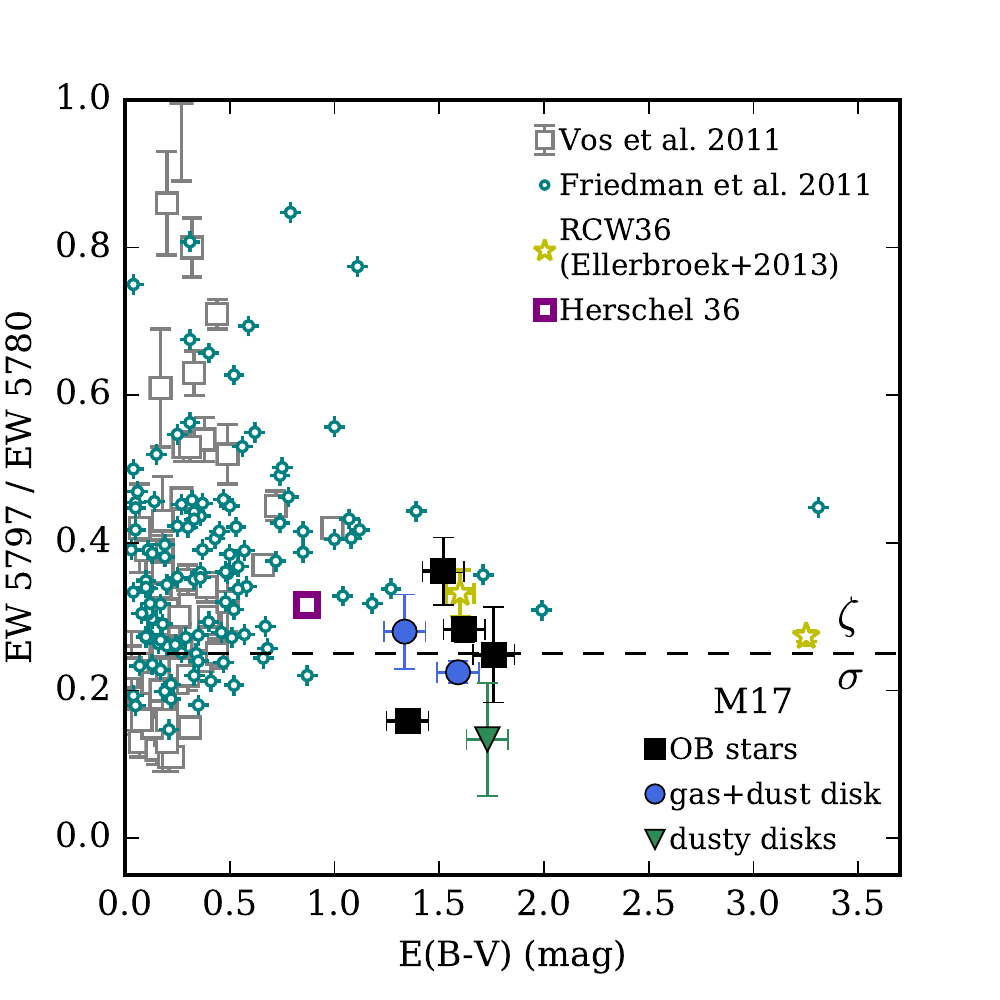}
    \caption{The ratio of the 5797/5780 DIBs. The open symbols show data from the literature and the filled symbols data from the sightlines towards M17. The dashed line represents the division between $\zeta$ and $\sigma$ type clouds.}
    \label{P3:fig:5797/5780EBV}
\end{figure}

\citet{2011A&A...533A.129V} demonstrated that the ratio 5797/5780 might be useful to distinguish between sight lines probing diffuse cloud edges and denser cloud cores. The 5797~\AA\ DIB is less sensitive to UV radiation than the 5780~\AA\ DIB, therefore, one would expect that the 5797/5780 ratio is lower in higher extinction environments.
This effect is strongest in sightlines dominated by one velocity component in the interstellar gas absorption profile (i.e. single-cloud sightlines) and tends to get 'washed-out' in multi-cloud sightlines.
The EW ratio of the 5797/5780 DIBs is plotted in Fig.~\ref{P3:fig:5797/5780EBV}. Compared to the lower extinction data (E(B-V) $<$ 1 mag; \citealt{2011ApJ...727...33F}; \citealt{2011A&A...533A.129V}) the DIBs towards M17 cover a smaller range (0.15-0.45) in the 5797/5780 DIB ratio (though at higher extinction values).
We do not find a clear correlation with spatial location in the cloud.
Note that the interstellar gas absorption features of \ion{Ca}{ii} and \ion{Na}{i} are largely unresolved at the spectral resolution of X-shooter.
It is conceivable that at higher spectral resolution the interstellar gas component associated with M17 will be resolved in multiple velocity components. This could imply that due to averaging along the line-of-sight the observed 5797/5780 ratio constitutes an average of this ratio in individual clouds.
Variations of 5797/5780 (in diffuse ISM) are thought to depend strongly on the local geometry and the effective UV radiation field (see above), both of which are complex within the large \ion{H}{ii} region. Additionally, variations of the 5797/5780 ratio in the densest regions (high $R_V$) can potentially be driven by different mechanisms (e.g. depletion) as compared to changes in this ratio in the more diffuse medium (e.g. effective UV radiation field); this could be the case for M17 as the change on strength (per $A_V$) of these two DIBs with $R_V$ is different from each other as shown in Fig.~\ref{P3:fig:EWAV_vs_Rv} and discussed in  Sect.~\ref{P3:sec:DIBvsRv}. 

\subsection{8620 DIB}

The DIB at 8620~\AA\ has recently gained more attention because it has been included in the large-scale surveys of RAVE and \emph{Gaia}. It has been shown that the EW of this DIB is well correlated with the colour excess $E(B-V)$; \citet{2008A&A...488..969M} reported a relation between the DIB strength and $E(B-V)$ given by $E(B-V)=(2.72\pm0.03)\times{\rm EW~8620~\AA}$ for $E(B-V)<1.2$. 

\citet{2015A&A...583A.132M} measured this DIB in two sightlines with high extinction towards O stars in the cluster Berkeley~90. The authors detect at least two clouds moving at different velocities towards these objects; one thinner cloud, associated with the material far away from the stellar cluster, and a thicker one, associated with the material local to the cluster. The latter cloud is thicker for one sight line, which allowed the authors to measure the properties of DIBs in $\sigma$ ($R_V\sim4.5$) and in $\zeta$ ($R_V\sim3.1$) clouds (see Sec.~\ref{P3:sec:ratio_5797_5780} for details about $\sigma$ and $\zeta$ clouds). They observed that the DIB at 8620~\AA\ is highly depleted in dense clouds and, therefore, correlates poorly with extinction higher than $A_V\sim6$.  

\citet{2016MNRAS.463.2653D} extended
the correlation of the 8620 DIB with extinction by adding sight lines with higher values of $A_{K}$. They used the extinction in the $K$ band because it is less sensitive to the grain size than $A_{V}$. They adopted $A_{Ks}=0.29E(B-V)$, which allowed to express the \citet{2008A&A...488..969M} relation as $A_{Ks}=(0.691)\times{\rm EW~8620~\AA}$. They find the correlation:
\begin{equation}
    A_{Ks} = (0.612 \pm 0.013)\ {\rm EW} + (0.191 \pm 0.025)\ {\rm EW^{2}} \, .
\label{P3:eq:Dami}
\end{equation}

This is in good agreement with \citet{2008A&A...488..969M} for ${\rm EW < 0.6}$~\AA\ which translates to $A_V<9$~mag for the extinction properties of their sample.

\begin{figure}
    \centering
    \includegraphics[width=0.98\hsize]{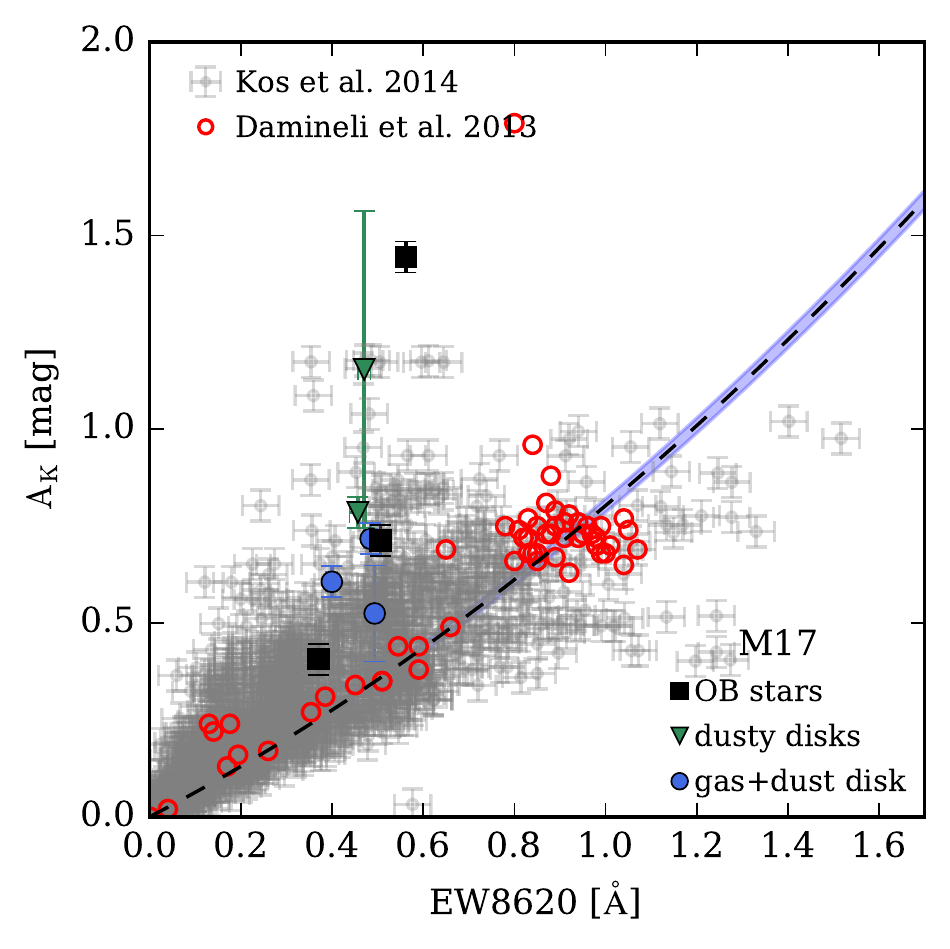}
    \caption{$K$-band extinction $A_K$ versus strength of the DIB at 8620~\AA. The open symbols show data from the literature and the filled ones sightlines towards M17. The dotted line and the blue area show the relation in Eq.~\ref{P3:eq:Dami}.
    }
    \label{P3:fig:DIB8620_vs_Ak}
\end{figure}

Using data from the RAVE survey, \citet{2014Sci...345..791K} measured the strength of the 8620~\AA\ DIB for $\sim$500\,000 spectra at distances $<3$~kpc. We plot the EW of the 8620~\AA\ DIB in Fig.~\ref{P3:fig:DIB8620_vs_Ak}. We include the data by \citet[red dots]{2016MNRAS.463.2653D} and \citet[grey dots]{2014Sci...345..791K}, the black-dashed line and the blue shadow represent the relation in Eq.~\ref{P3:eq:Dami}. In order to derive $A_K$ from the \citet{2014Sci...345..791K} data, we adopted $A_K/A_V=0.11$.

The DIB strength towards M17 shows a very large spread in comparison with the relation by \citet{2016MNRAS.463.2653D}. This spread is similar to that observed in the sight lines from the RAVE survey. The large scatter in the M17 data could be an indication that the DIB carrier gets depleted at extinctions $A_K > 1.0$. We detect a general positive correlation between the strength of the 8620 DIB and the $K$-band extinction for $A_K \lesssim 1.0$. 

\section{DIB strengths versus $R_V$}
\label{P3:sec:DIBvsRv}

\begin{figure}
     \centering
     \includegraphics[width=\hsize]{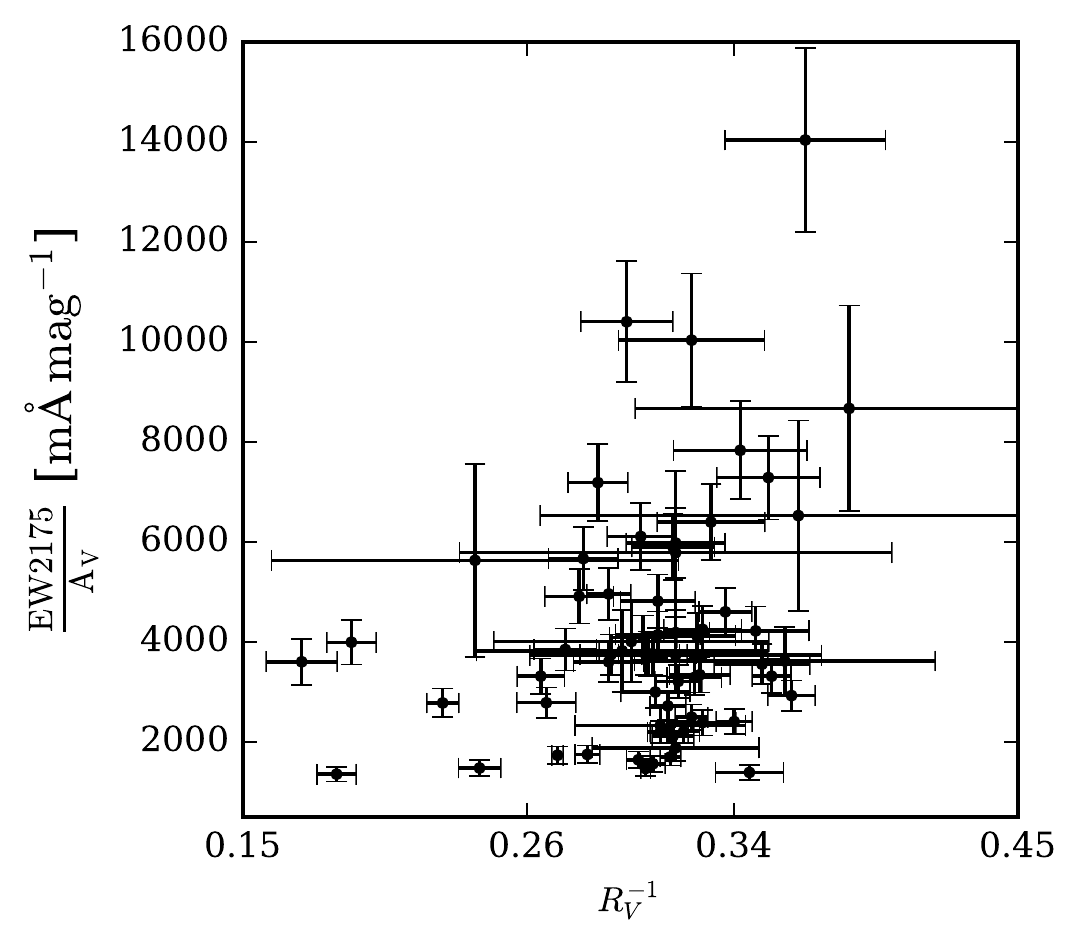}
     \centering
     \caption{Normalised EW of the 2175~\AA\ feature measured in Galactic sight lines as a function of $R_{V}$. The measurements are obtained from \citet{2017ApJ...835..107X}. The 2175~\AA\ feature is stronger in sight lines characterised by a smaller value of $R_{V}$.}
     \label{P3:fig:2175}
\end{figure}

The value of $R_V$ is related to the size distribution of the dust grains. Large values of $R_V$ imply a relative over-abundance of large grains, whereas small $R_V$ values indicate that the small dust grains are more dominant in the size distribution. Typically higher values of $R_V$ are associated with denser clouds where grain growth is or has been  more significant than in diffuse regions. This is reflected in the relative slope of the extinction curve. Larger $R_V$ values correspond to a flatter extinction curve, meaning that photons at all wavelengths get more equally absorbed. 

\citet{1989ApJ...345..245C} presented observational evidence that the 2175~\AA\ feature, normalised by the amount of extinction, shows a positive trend with $R_{V}^{-1}$. In Fig.~\ref{P3:fig:2175} we present a similar plot, based on measurements recently collected by \citet{2017ApJ...835..107X}. Although the plot reveals significant scatter, there appears to be a general trend that sight lines with a smaller value of $R_{V}$ show a stronger 2175~\AA\ feature per magnitude visual extinction. 
The Pearson correlation coefficient between EW(2175\AA) and $R_V^{-1}$ has a value of $r = 0.3$, which indicates a rather weak formal correlation.

\begin{figure*}%[th!]
    \centering
    \includegraphics[width=0.93\hsize]{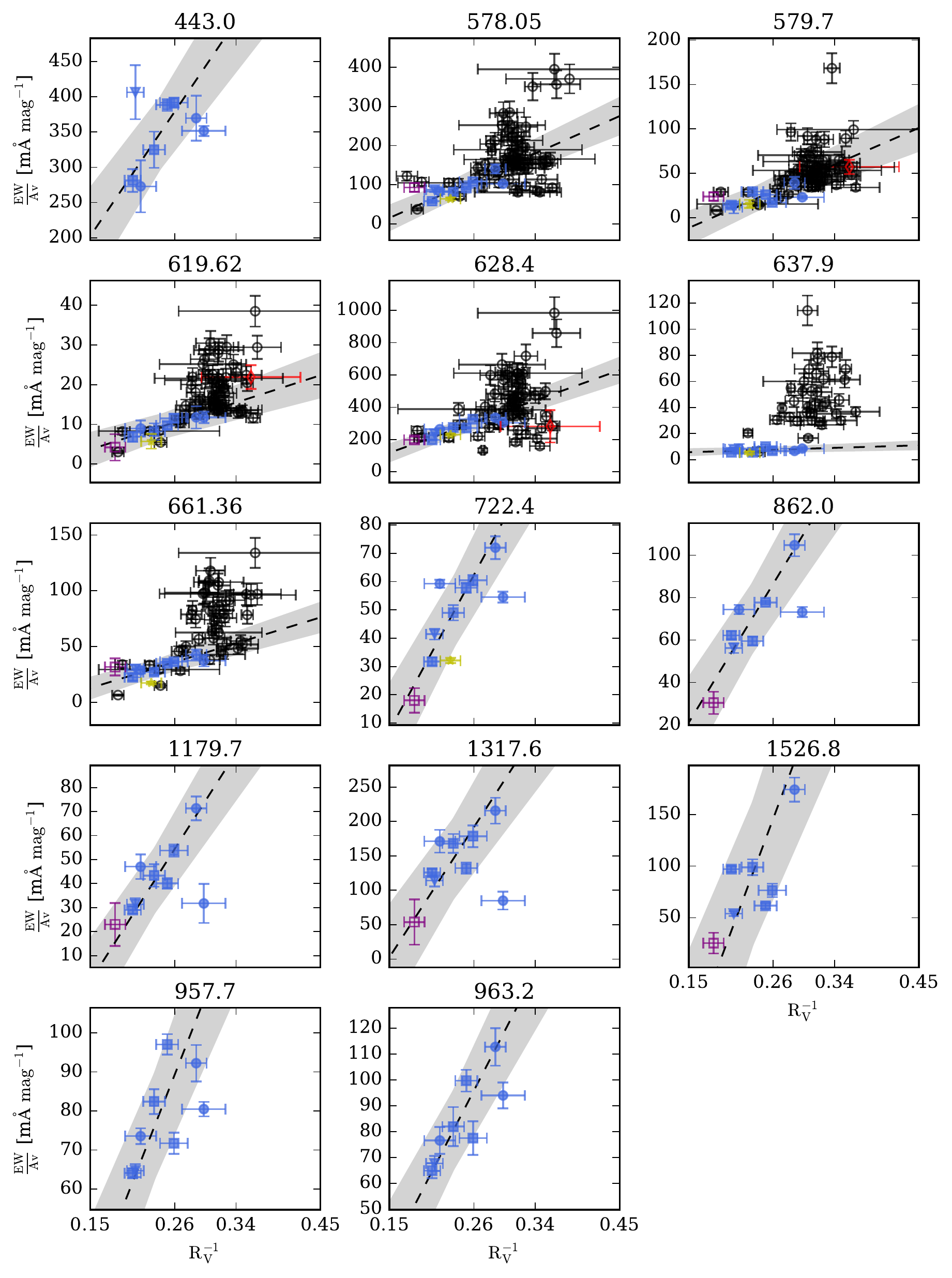}
    \caption{Normalised DIB EW versus $R_V^{-1}$. The blue symbols show the M17 data; the dots show the stars in M17 with circumstellar disks, the triangles the stars with an IR excess longwards of 3~$\mu$m, and the squares the OB stars in M17. The open symbols show data from the literature. The purple square shows the position of Herschel\,36, the yellow star shows the source 408 in RCW\,36, and the red diamond shows the high mass X-ray binary 4U1907+09. The grey dots show the data by \citet{2017ApJ...835..107X} when available. The black line shows a linear fit with $R_V^{-1}$ and the shaded region the 1-$\sigma$ error box of the fit to the M17 measurements only.
    }
    \label{P3:fig:EWAV_vs_Rv}
\end{figure*}

In Fig.~\ref{P3:fig:EWAV_vs_Rv} we plot the strength of the DIBs normalised by the visual extinction $A_V$ with $R_V^{-1}$. We include the sight lines towards M17, Herschel\,36, and the object 408 in RCW\,36. Where available, we add the data from \citet{2017ApJ...835..107X} who compiled the EW of several DIBs in 97 sight lines. For most DIBs we discern a positive trend of the normalised DIB strength (EW/$A_V$) with $R_{V}^{-1}$, similar to the trend seen for the 2175~\AA\ feature (Fig.~\ref{P3:fig:2175}). Given that $A_V$ is in the denominator of both axes, if the uncertainties in $A_V$ are large, a spurious relation between EW/$A_V$ and $R_V^{-1}$ might result. This is because when dividing by a very uncertain $A_V$, one could expect some values of both EW/$A_V$ and $R_V^{-1}$ to be very low and some to be very large causing a spurious linear relation intercepting at zero. However, we consider the errors in $A_V$ are small enough -- primarily due to the accurate spectral type determination -- that we do not expect such an effect here.

We note that there is a cloud of data points centred at the Galactic average value of $R_V = 3.1$. The observed scatter could represent sampling of non-heterogeneous (multiple) interstellar cloud conditions in these sightlines. In other words, variations in DIB band strength (per unit reddening) are more pronounced than variations in \emph{average} dust grain size. The presence of this cloud of data points shows that normalizing by $A_V$ does not introduce a spurious relation. If that would be the case, we would expect the points at all $R_V$ values to follow a linear relation. On the other hand, for sightlines with $R_V \gtrapprox 3.8$ $(R_V^{-1} \lessapprox 0.26)$ there appears to be a much more pronounced (linear) relation between EW/$A_V$ and $R_V^{-1}$. The M17 data presented here strengthen the presence of this linear relation that can already be gleaned from the data compiled by \citet{2017ApJ...835..107X} where most of the sightlines with $R_V \gtrapprox 3.8$ correspond to stars in star forming regions (Orion giant molecular cloud, Upper Scorpius, Triffid nebula, and Rho Ophiuchi).

We performed an orthogonal distance regression \citep[ODR,][]{1990odr_reference}\footnote{Information about the python package is available at \url{https://docs.scipy.org/doc/scipy/reference/odr.html}} to fit a linear function to the DIB strength normalised by the visual extinction $A_V$ versus $R_V^{-1}$ for sightlines towards M17. The fit results, together with the Pearson correlation coefficients are listed in Table~\ref{P3:tab:fitresults} and shown in Fig.~\ref{P3:fig:EWAV_vs_Rv} where the black line corresponds to the best fit to the M17 data alone and the shaded region to the 1$\sigma$ error on the fit.

\begin{table}
\caption{Results of the linear fit of EW/$A_V$ versus $R_V^{-1}$ towards M17 sightlines.} \label{P3:tab:fitresults}
\centering
\begin{tabular}{cccc}
\hline\hline
DIB & slope & interception & Pearson r\\
 & [${\rm m\AA/mag}$] & [${\rm m\AA/mag}$] & \\
\hline
4430 & $1604\pm403$ & $-38\pm97$ & 0.39 \\
5780 & $868\pm195$ & $-115\pm45$ & 0.80 \\
5797 & $375\pm110$ & $-67\pm26$ & 0.63 \\
6196 & $62\pm24$ & $-5\pm5$ & 0.91 \\
6284 & $1705\pm324$ & $-137\pm76$ & 0.85 \\
6379 & $17\pm13$ & $3\pm3$ & 0.21 \\
6614 & $211\pm56$ & $-18\pm13$ & 0.88 \\
7224 & $499\pm99$ & $-67\pm23$ & 0.70 \\
8620 & $594\pm130$ & $-68\pm30$ & 0.69 \\
11797 & $500\pm106$ & $-75\pm24$ & 0.46 \\
13176 & $1714\pm462$ & $-254\pm107$ & 0.11 \\
15268 & $1992\pm551$ & $-372\pm128$ & 0.63 \\
9577 & $501\pm107$ & $-41\pm25$ & 0.62 \\
9632 & $573\pm121$ & $-53\pm28$ & 0.82 \\

\hline
\end{tabular}
\end{table}

Based on the results from the fit, we confirm that for M17 there is a linear relation between the normalised DIB strength and $R_V^{-1}$. This correlation is strongest for the DIBs at 5780, 6196, 6284, 6614, and 9632~\AA. From the Pearson correlation coefficient, $r$, we find that the correlation is moderate to strong ($r>0.6$) for all DIBs except for those at 4430, 6379, 11797, and 13176~\AA\ where the correlation is weak ($r<0.5$). 
We note that for the 11797 and 13176~\AA\ DIBs there is a clear outlier data point which strongly affects the Pearson correlation coefficient $r$ (Fig.~\ref{P3:fig:EWAV_vs_Rv}). 
This point corresponds to the B275 sight line where the spectral quality in the ranges of these two DIBs is poor, which could introduce systematic errors in the DIB measurements (see Figs.~\ref{P3:fig:ap11797} and \ref{P3:fig:ap13176}).  
Excluding this point results in $r>0.7$ for both DIBs.

\section{Discussion and conclusion}
\label{P3:sec:discussion}

\subsection{Extinction in M17}

We calculated the extinction properties of the sight lines towards M17, with overall relatively large $R_V$ values, using different methods; \emph{i)} by dereddening the SED of the stars using Cardelli's extinction law and then fitting it to Kurucz models to find $A_V$ and $R_V$ (from that $E(B-V)=A_V/R_V$). \emph{ii)} Directly by comparing the intrinsic magnitudes to the observed ones, to find $A_V$ and $E(B-V)$ ($R_V=A_V/E(B-V)$). And \emph{iii)} by finding $A_V$ and $E(B-V)$ from the intrinsic and observed magnitudes and calculating $R_V$ as defined by \citet[Eq.~\ref{P3:eq:Rv_fitzpatrick}]{2007ApJ...663..320F}. We observe that the colour excess $E(B-V)$ calculated with Cardelli's extinction law and the SED fitting is systematically higher than using the other two methods. This confirms that this extinction law is not well suited for large values of $R_V$ as already stated in \citet{1989ApJ...345..245C}. We conclude that the best way of obtaining the extinction properties is to use the intrinsic and observed magnitudes and then calculate $R_V$ as shown in Eq.~\ref{P3:eq:Rv_fitzpatrick}, because with this equation it is also possible to take into account the NIR part of the SED.

\subsection{Relation of DIBs with E(B-V)}

In Fig.~\ref{P3:fig:DIBvsEBV} we show the relation of the DIBs at 5780 and 5797~\AA\ with the colour excess $E(B-V)$ in M17 and compare it with other sight lines published in the literature. $E(B-V)$ for the sight lines in M17 is relatively high compared to other Galactic sight lines; nevertheless, the DIB strength seems to follow the same trend as observed in other reddened sight lines. 

\citet{2013A&A...551A...5E} reported an anomalous behaviour of the DIB strength in sight lines towards the star forming region RCW\,36. We do not find a similar deviation in the sight lines towards M17. The behaviour reported in RCW\,36 could be due to the way in which the authors calculated $A_V$. They assumed a value of $R_V=3.1$, dereddened the spectrum using Cardelli's extinction law, and then fitted it to Kurucz models. In this paper we show that the value of $E(B-V)$ can be overestimated by as much as 0.5~mag (Fig.~\ref{P3:fig:E_BV_comparison}) when using Cardelli's extinction law. Previous studies have shown that star-forming regions such as M17 and M8 (where Herschel~36 resides) often show high values of $R_V$ \citep[e.g.][]{1989ApJ...345..245C,1997ApJ...489..698H}. 
Therefore the assumption that $R_V=3.1$ for RCW\,36 probably underestimates its true $R_V$.

\subsection{Relation of DIBs with $R_V$}

The observed trend of EW(DIB)/$A_V$ with $R_V^{-1}$ for M17 and other sightlines with $R_V \gtrapprox 3.8$  ($R_V^{-1} \lessapprox 0.26$; cf. Figure~\ref{P3:fig:EWAV_vs_Rv}) is reminiscent to what \citet{1989ApJ...345..245C} and \citet{2017ApJ...835..107X} find for the 2175~\AA\ bump, the carrier of which is thought to consist of some (unidentified) carbonaceous material; however, the trend of the bump is weaker than that observed for the DIBs.

The slope of the linear relation between EW/$A_V$ and $R_V^{-1}$ (valid for $R_V \gtrsim 3.8$) can be interpreted as a measure of the sensitivity to which variations in the abundance of DIB carriers (per unit visual extinction) are related to variations in the average grain size of interstellar dust. 
This slope can then be interpreted in the context of either the depletion of the DIB carrier as the dust grains coagulate and grow \citep[see e.g.][EW/$A_V$ decreases while $R_V$ increases]{1980ApJ...235...63J,2009A&A...502..845O} or the production of DIB carriers as the larger grains get destroyed by the increasingly effective UV radiation field \citep[see e.g.][EW/$A_V$ increases and $R_V$ decreases]{1995ApJS..100..187C,2004ASPC..309..347J}.

Alternatively, in the hypothesis that some DIB carriers are ionized species, the strengths of DIBs are expected to increase with a stronger effective UV radiation field. If $R_V$ could be regarded as a tracer of the effective UV radiation field, higher values correspond to a lower ionization fraction, hence lower EW/$A_V$. 
In this context the slope of EW/$A_V$ versus $R_V^{-1}$ would be directly related to the ionization potential of the carrier species.

\begin{figure*}[ht!]
    \centering
    \includegraphics[width=0.9\hsize]{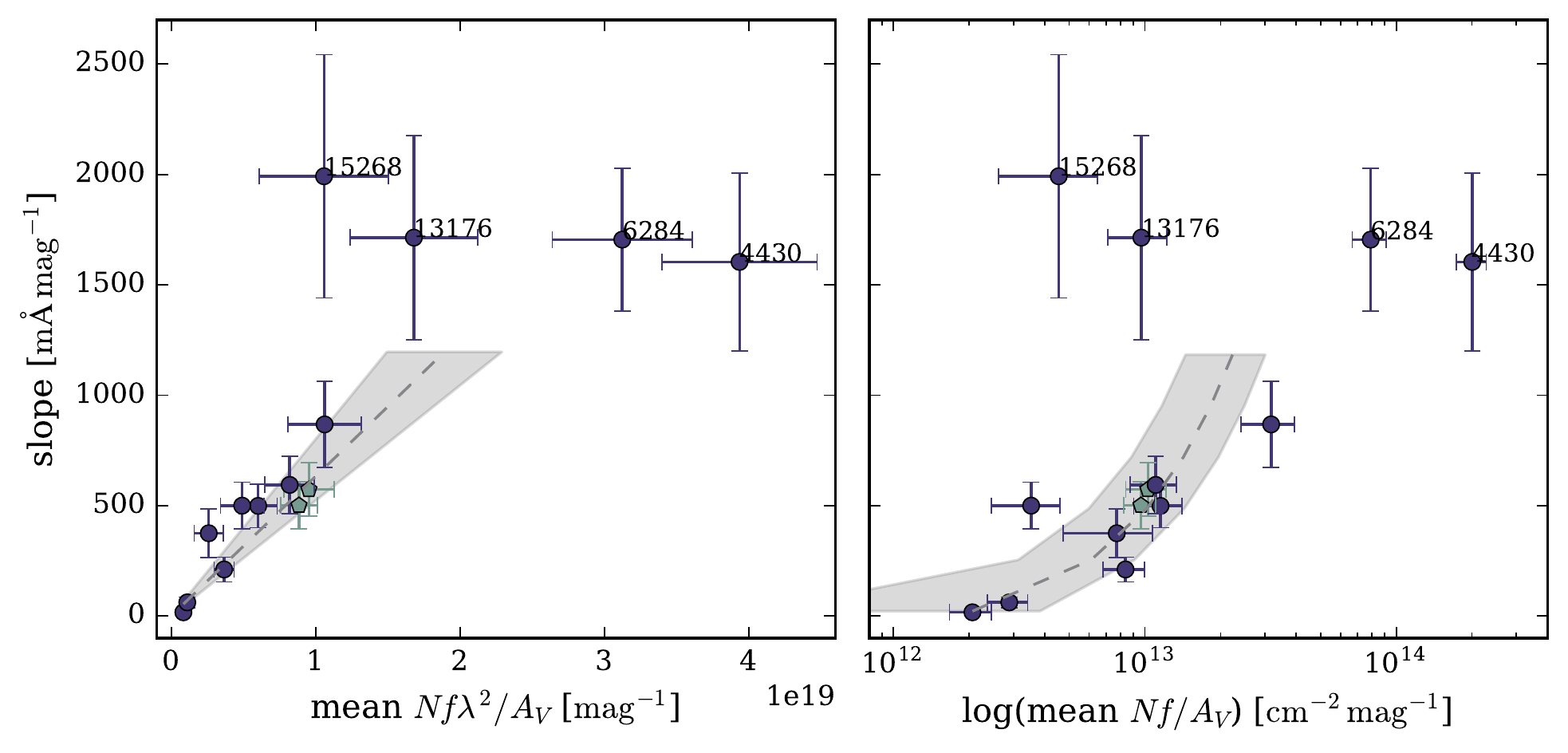}
    \caption{Slope representing the sensitivity of EW/$A_V$ with changes in $R_V^{-1}$ versus mean $A_V$ normalised line strength (left; expressed as $Nf/A_V$) and versus the log of this quantity (right) for all sightlines. The dashed line and the shaded regions represent the linear fit to the group of DIBs with slopes $\lesssim 1000$ and the 2$\sigma$ error bars to the fit. The two $C_{60}^{+}$ DIBs are represented with light green pentagons.}
    \label{P3:fig:strength_vs_slope}
\end{figure*}

With the current data we can not distinguish between these different scenarios.
However, it is possible to identify groups or families of DIB carriers which behave similarly in the transition from the more dense to diffuse ISM (high to low $R_V$). 
To investigate this we compared the column density of the DIB carriers multiplied by the (unknown) oscillator strength, $Nf$, with the slope d(EW/$A_V$)/d$R_V^{-1}$ (cf. Fig.~\ref{P3:fig:EWAV_vs_Rv}; Table~\ref{P3:tab:fitresults}). This slope can be viewed as a measure of the sensitivity of the DIB carriers to changes in $R_V^{-1}$, hence in grain size distribution or as a measure of the carrier's ionization potential. We will refer to this slope as the DIB-$R_V$ sensitivity.
Assuming that the optically-thin approximation holds for the DIBs (no saturation of absorption) we have:
\begin{equation}
Nf = 1.13 \times 10^{20} \,\frac{\mathrm{EW}}{\lambda^{2}}
\end{equation}
where $N$ is the column density in ${\rm cm^{-2}}$, $f$ the oscillator strength, EW is the equivalent width (in \AA) and $\lambda$ is the central wavelength of each DIB (in \AA).
We show the dependence of the DIB-$R_V$ sensitivity on $Nf\lambda^2$, and $Nf$ in Fig.~\ref{P3:fig:strength_vs_slope}; 
In both cases a linear fit is shown with a dashed line and the shaded region shows the 2$\sigma$ error to the fit. Additionally we marked the $C_{60}^{+}$ DIBs with green pentagons.

First we discuss the results in the context of $R_V$ as a proxy for changes in the grain size distribution.

We are able to identify two groups; the first group contains ten moderately strong DIBs, including the two $C_{60}^+$ DIBs, the DIBs presenting a red wing in Herschel~36 (5780, 5797, and 6614), and those at  6196, 6379, 7224, 8620, and 11797~\AA. These DIBs show a modest sensitivity to the change in grain size, and their sensitivity increases about linearly as a function of DIB strength. 
The second group consists of four strong DIBs (at 4430, 6284, 13176, and 15268~\AA), that are very sensitive to the change in grain size distribution (large d(EW/$A_V$)/d$R_V^{-1}$ slope values), but their DIB-$R_V$ sensitivity does not seem to depend on the DIB strength.

In the case of the first group, we postulate that the observed linear increase of the DIB-$R_V$ sensitivity with the amount of DIB carriers, $Nf$, suggests that the increase of the latter is primarily due to an increase in column density and that $f$ is quite similar for this group of DIB carriers. 
This is because the reverse situation, i.e that a systematic increase of the sensitivity of DIB carriers to changes in $R_V$ is related to a systematic increase of the oscillator strength of the DIB carriers appears much more unlikely. 
If this is the case, the oscillator strengths of these DIBs could be extracted with reference to the two DIBs at 9577 and 9632~\AA\ assigned to C$_{60}^+$ ($f_{9577}=0.018$ and $f_{9632}=0.015$).

The linear dependency of the DIB-$R_V$ sensitivity on the DIB strength implies that the relative production/destruction rate of these DIB carriers is the same for the same change in typical grain size ($R_V$). In other words, for a doubling of the abundance of DIB carriers in the first group, the same change in grain size distribution is required. This may imply a common production/destruction mechanism. The fact that the DIB-$R_V$ sensitivity of the stronger DIBs is independent of their strength suggests the existence of an additional reservoir of DIB carriers unrelated to the dust grains. These carriers would be produced/destructed as the grain size distribution changes as well as via other mechanisms. 

Alternatively, we consider the scenario where $R_V$ is a proxy of the effective UV radiation field strength and the DIB carriers are large ionised molecules. 
In this case Fig.~\ref{P3:fig:strength_vs_slope} has to be interpreted differently. Here the slope d(EW/$A_V$)/d$R_V^{-1}$ is an indirect measure of the carrier's ionization potential. 
The broad DIBs at 15268, 13176, 6284 and 4430~\AA\ have a (similar) strong slope which implies a (similar) low ionization potential. The ionization potential of the carriers of these broad DIBs is then expected to be significantly lower than the ionization potential of 7.58~eV measured for $C_{60}$ (\citealt{DEVRIES1992159}). In the opposite direction, the narrow DIBs at 7224, 5797, 6614, 6196, and 6379~\AA\ which have increasingly shallower slopes should then have increasingly higher ionization potentials.
For PAHs the ionisation potential is strongly related to their number of $\pi$-electrons (see Fig.~B.1. in \citealt{2005A&A...432..515R}).

We note that in the case that there are several foreground clouds contributing to the extinction in the sightline towards M17, the interpretation of the observed relations is complicated. The hypotheses presented above assume that most of the variation in $A_V$ and $R_V$ arises from the dust and gas in the M17 region. Correcting for a foreground sheet of dust, with $A_V = 2$ and a constant foreground EW(DIB)/$A_V$ contribution projected across the M17 region, would shift all the EW/$A_V$ values vertically.
Also, correcting for foreground dust with a nominal $R_V$ = 3.1, will sightly increase the $R_V$ values derived for the local M17 dust, but have little impact on the slope of the relation between EW(DIB)/$A_V$ and $R_V^{-1}$. It could also affect the mean normalised line strengths. The above scenarios and hypotheses can be further tested by examining the relation between EW(DIB)/$A_V$ and $R_V$ for other star forming regions that probe a range of extinction properties and grain size distributions, similar to M17.
This will also help to disentangle the impact of foreground dust extinction and to determine possible effects due to measurement uncertainties.

\section{Summary}
\label{P3:sec:summary}

We present an analysis of the properties of the DIBs in sight lines toward the star forming region M17 in comparison to the properties of nominal DIBs as well as DIBs that have been reported to be anomalous.

\begin{enumerate}
    
    \item Cardelli's extinction law is not suitable for large values of $R_V$. The best way of calculating the extinction parameters $A_V$ and $E(B-V)$, when the spectral type of the observed star is known, is by directly comparing the observed and intrinsic magnitudes. By using the relation between $R_V$ and $E(V-K)/E(B-V)$ of \citet{2007ApJ...663..320F} it is possible to calculate $A_K$ and, therefore the $K$-band excess produced by the circumstellar disk. 
    
    \item From the observed C$_{60}^+$ DIB at 9577~\AA\ we derive a column density N(C$_{60}^+$)$= 2.5-4.3\times 10^{13}$~cm$^{-2}$. 
    
    \item We also detect the NIR DIBs at 11797, 13176, and 15268~\AA\ in the sightlines towards M17.
    
    \item The profiles of the DIBs toward M17 do not present a red wing, in contrast to those towards Herschel\,36. This indicates that the red wing is not common to star forming regions, but that the environment towards Herschel\,36 must present some special properties that cause the DIBs to have a red wing.
    
    \item The strength of the 5780~\AA\ DIBs towards M17 is as expected from the relations found in the literature for their values of $E(B-V)$. The 5797~\AA\ DIBs show a relatively large spread in relation to other Galactic DIBs, and the 6614~\AA\ DIBs towards M17 are slightly weaker than expected for their $E(B-V)$ values.

    \item In the M17 region we find trends between the strength of the studied DIBs (per unit visual extinction) and $R_V^{-1}$, most notably for the 6196 and 7224~\AA\ DIBs. The trend remains when we include the sightlines towards star forming regions presented in \citet{2017ApJ...835..107X}. For sightlines with $R_V\sim3.1$ there is no trend. 
    
    \item The slope values of the linear relation between the DIBs EW normalised by $A_V$ and $R_V^{-1}$ (DIB-$R_V$ sensitivity) allow us to identify two main groups of DIBs in terms of their sensitivity to $R_V$ (for $R_V > 3.8$, $R_V^{-1} < 0.26$). The first group consists of DIBs with similar sensitivity to changes in $R_V$ as the two $C_{60}^{+}$ DIBs: 5780, 5797, 7224, 8620, 9577, 9632, and 11797~\AA. For this group we find that the DIB-$R_V$ sensitivity varies linearly with DIB strength.
    The second group consists of DIBs with roughly three times stronger DIB-$R_V$ sensitivity: 4430, 6284, 13176, and 15268~\AA; for this group, the DIB-$R_V$ sensitivity does not depend on the DIB strength.
    This response of DIB carriers to changes in interstellar dust properties provides additional clues to the (dis)similarity of different DIB carriers, and may provide clues as to their production/destruction mechanisms, and can thus ultimately help with their identifications.
    
    \item Three scenarios are proposed for the observed behaviour of DIBs in star-forming regions (with $R_V > 3.8$):
    (1) DIB carriers stick to dust grains, and thus get depleted from the gas phase, as the grains grow in size (dust-coagulation) in the denser regions of interstellar clouds;
    (2) DIB carriers are produced from the dust grains in the strong UV radiation fields in these regions which lead to increased abundances of DIB carriers (and that of the 2715~\AA\ band) and, by direct consequence, a decrease of the average dust particle size.
    (3) As $R_V$ decreases and the effective UV radiation field strength increases, the ionization fraction of the parent carriers increases thus causing the DIBs related to ionized molecular carriers to become stronger.
    The effective response of DIB strength to changes in the UV radiation field, i.e. d(EW/$A_V$)/d$R_V^{-1}$, is then a measure of the carriers ionization potential. 
    At this point we can not distinguish between these scenarios. Further investigations of specific (star forming) regions that probe a range of $R_V$ values, like M17, are needed to examine the relation between DIB strength and $R_V$.
    
\end{enumerate}

\begin{acknowledgements}
Based on observations collected at the European Organisation for Astronomical Research in the Southern Hemisphere under ESO program(s) 091.C-0934(B) (Herschel\,36), 385.C-0720(A) (HD161056) and 60.A-9404(A), 085.D-0741, 089.C-0874(A), 091.C-0934(B) (M17). The authors thank Rens Waters, Martin Heemskerk, Juan Hernandez Santisteban, Lucia Klarmann, Samayra Straal, Marieke van Doesburgh, Xander Tielens, and Rosine Lallement for discussions that helped to improve this paper. 
This research made use of Astropy, a community-developed core Python package for Astronomy \citep{2018arXiv180102634T}, NASA's Astrophysics Data System Bibliographic Services (ADS), and the SIMBAD database, operated at CDS, Strasbourg, France \citep{2000A&AS..143....9W}.
\end{acknowledgements}

\bibliography{references}

\onecolumn
\begin{appendix}

\Online

 \section{Observed DIBs and gaussian fits}
 \label{P3:appendix_DIBs}

% ______________1_______________

\begin{figure*}[h]
    \centering
    \includegraphics[width=\hsize]{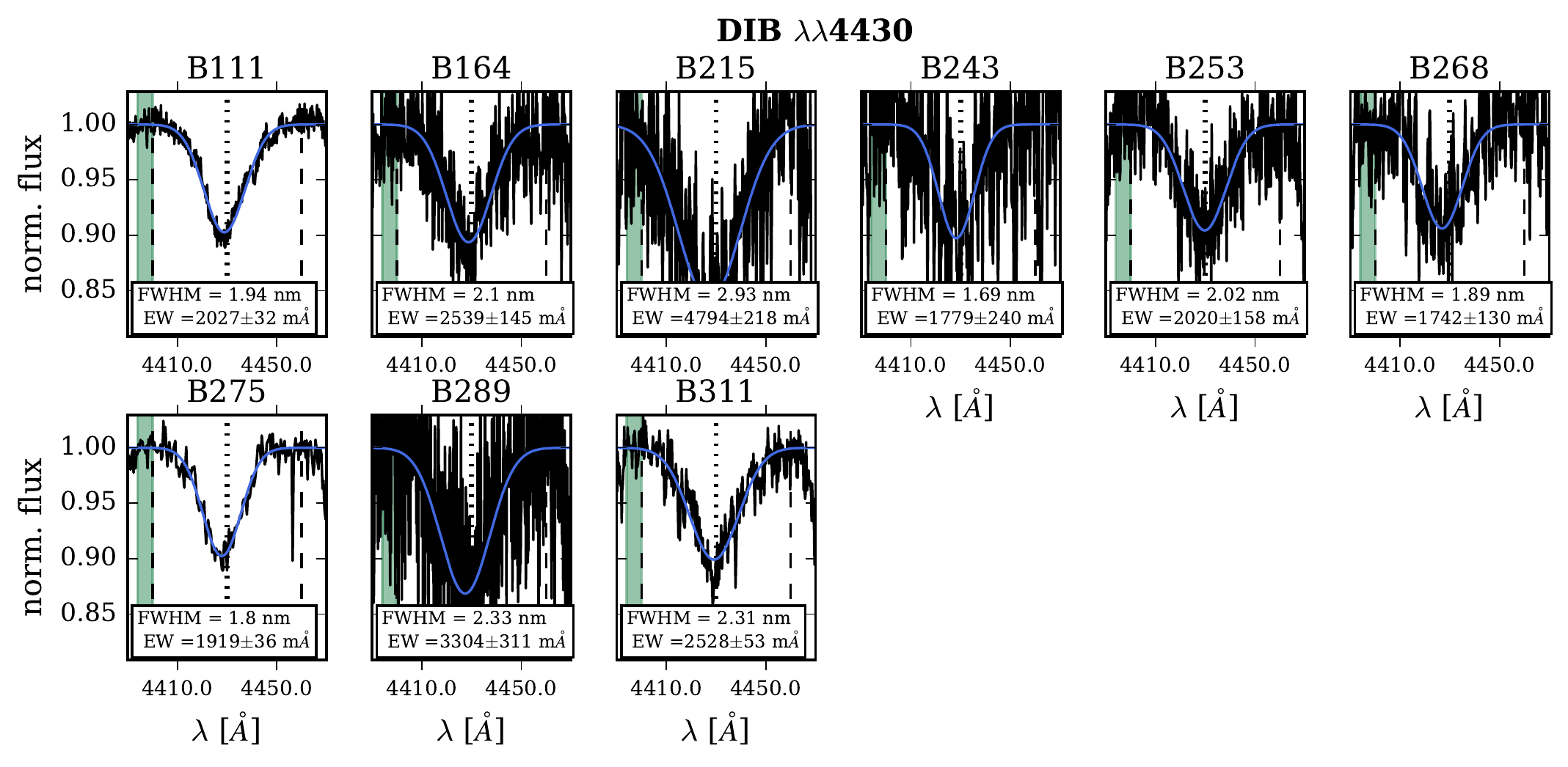}
    \caption{DIB profiles at 4430~\AA\ . The vertical dashed lines show the integration limits used to calculate the DIB strength, the dotted line shows the central wavelength of the DIB, and the green shaded region shows the region in which the error was calculated. With a solid-blue line we show the Gaussian fit to the DIB profile. The white box in the bottom of each panel indicated the FWHM and EW of this DIB for each object.}
    \label{P3:fig:ap4430}
\end{figure*}

% ______________2_______________

% ______________3_______________

\begin{figure*}
    \centering
    \includegraphics[width=\hsize]{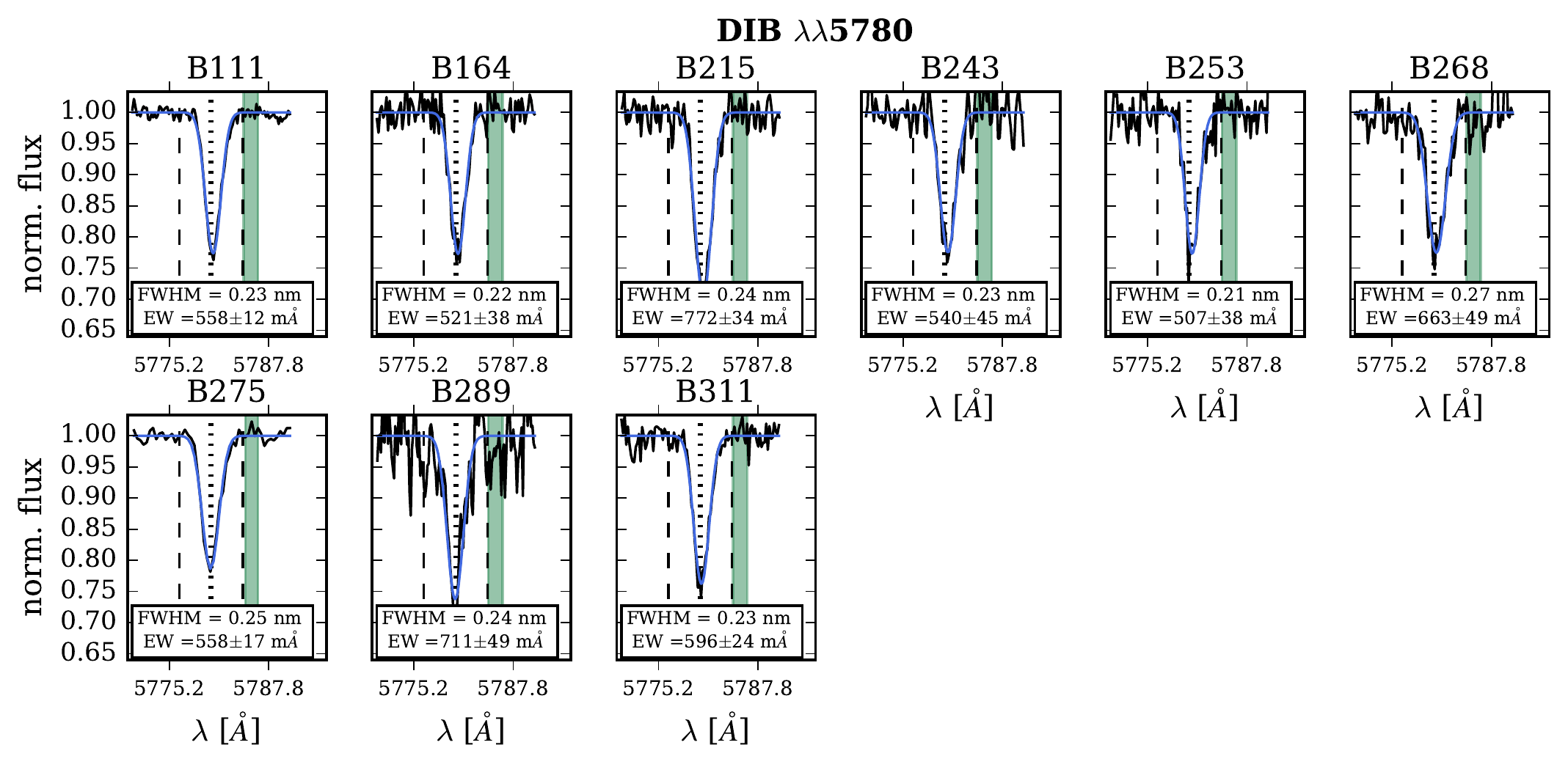}
    \caption{DIB profiles at 5780~\AA\ . The vertical dashed lines show the integration limits used to calculate the DIB strength, the dotted line shows the central wavelength of the DIB, and the green shaded region shows the region in which the error was calculated. With a solid-blue line we show the Gaussian fit to the DIB profile. The white box in the bottom of each panel indicated the FWHM and EW of this DIB for each object.}
    \label{P3:fig:ap5780}
\end{figure*}

% ______________4_______________

\begin{figure*}
    \centering
    \includegraphics[width=\hsize]{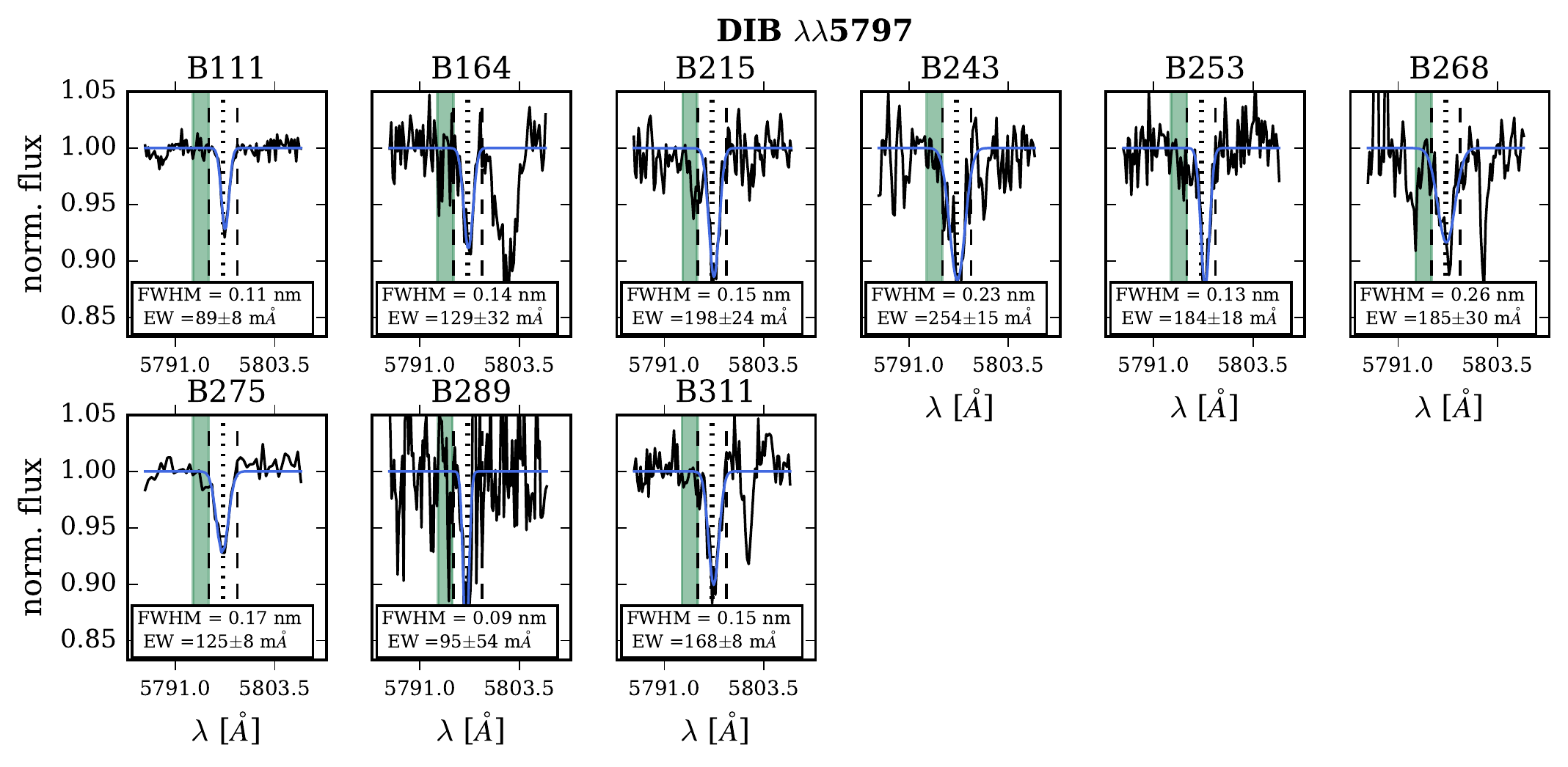}
    \caption{DIB profiles at 5798~\AA\ . The vertical dashed lines show the integration limits used to calculate the DIB strength, the dotted line shows the central wavelength of the DIB, and the green shaded region shows the region in which the error was calculated. With a solid-blue line we show the Gaussian fit to the DIB profile. The white box in the bottom of each panel indicated the FWHM and EW of this DIB for each object.}
    \label{P3:fig:ap5797}
\end{figure*}

% ______________5_______________

\begin{figure*}
    \centering
    \includegraphics[width=\hsize]{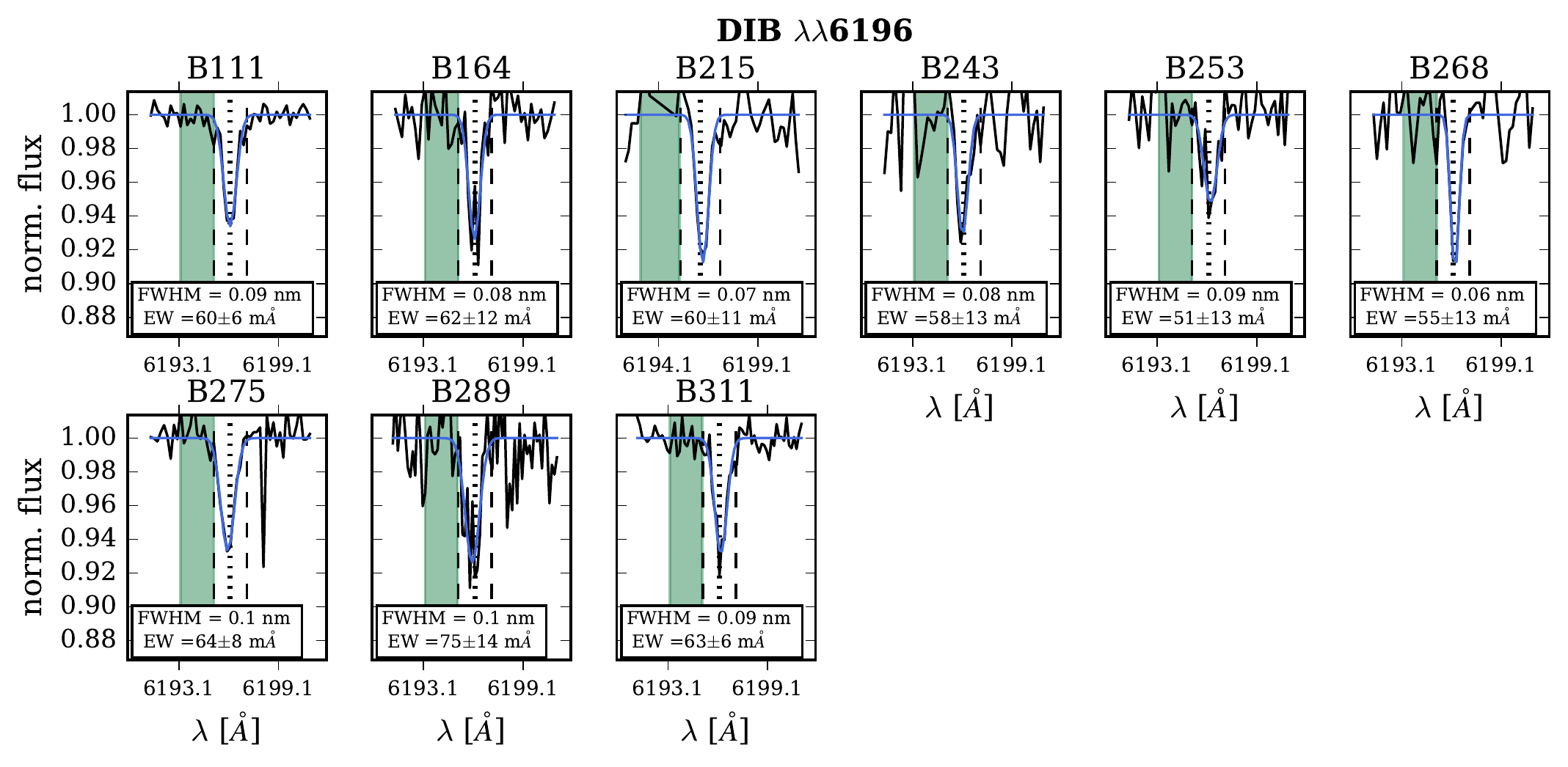}
    \caption{DIB profiles at 6196~\AA\ . The vertical dashed lines show the integration limits used to calculate the DIB strength, the dotted line shows the central wavelength of the DIB, and the green shaded region shows the region in which the error was calculated. With a solid-blue line we show the Gaussian fit to the DIB profile. The white box in the bottom of each panel indicated the FWHM and EW of this DIB for each object.}
    \label{P3:fig:ap6196}
\end{figure*}

% ______________6_______________

\begin{figure*}
    \centering
    \includegraphics[width=\hsize]{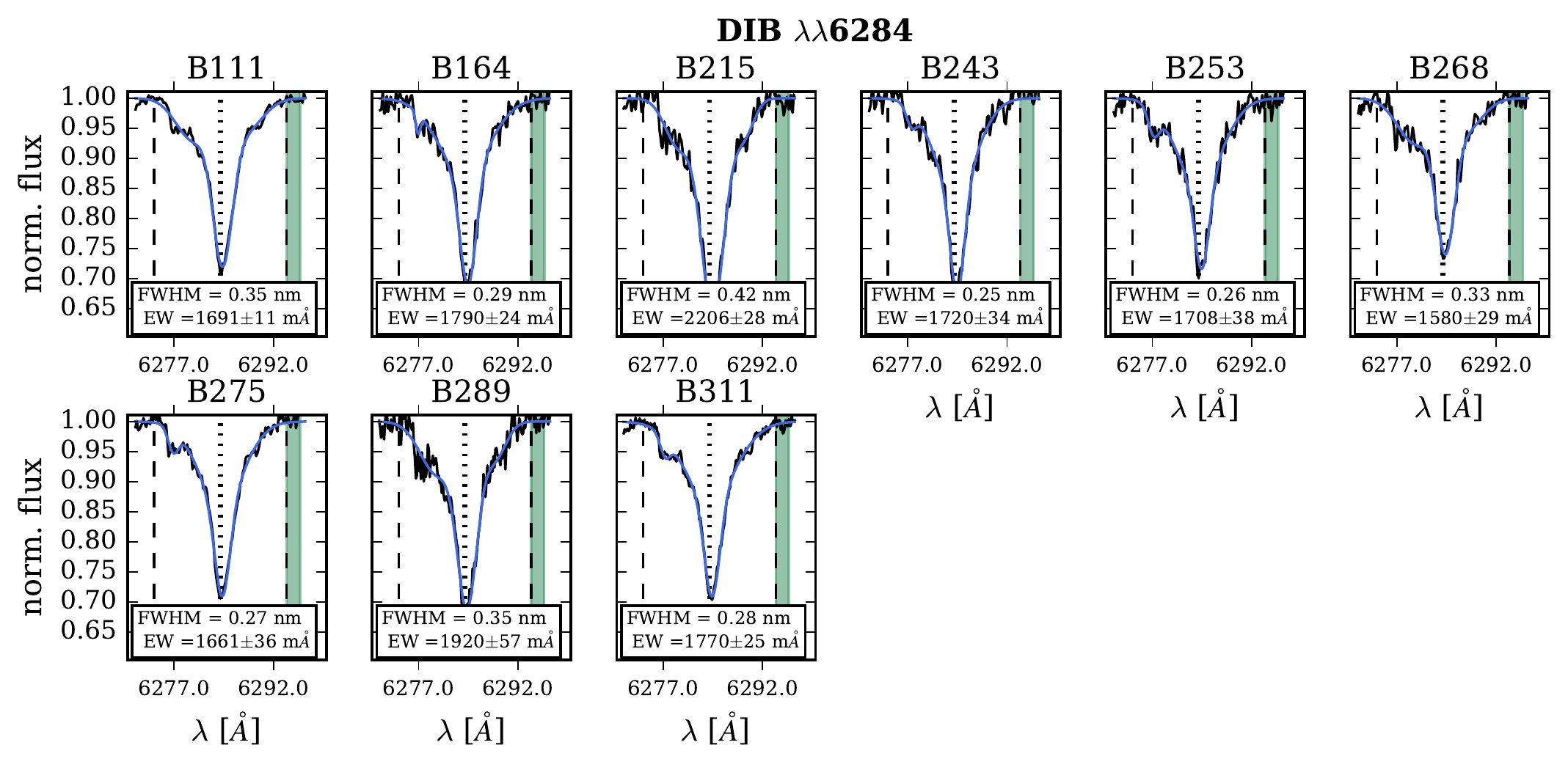}
    \caption{DIB profiles at 6284~\AA\ . The vertical dashed lines show the integration limits used to calculate the DIB strength, the dotted line shows the central wavelength of the DIB, and the green shaded region shows the region in which the error was calculated. With a solid-blue line we show the Gaussian fit to the DIB profile. The white box in the bottom of each panel indicated the FWHM and EW of this DIB for each object.}
    \label{P3:fig:ap6284}
\end{figure*}

% ______________7_______________

\begin{figure*}
    \centering
    \includegraphics[width=\hsize]{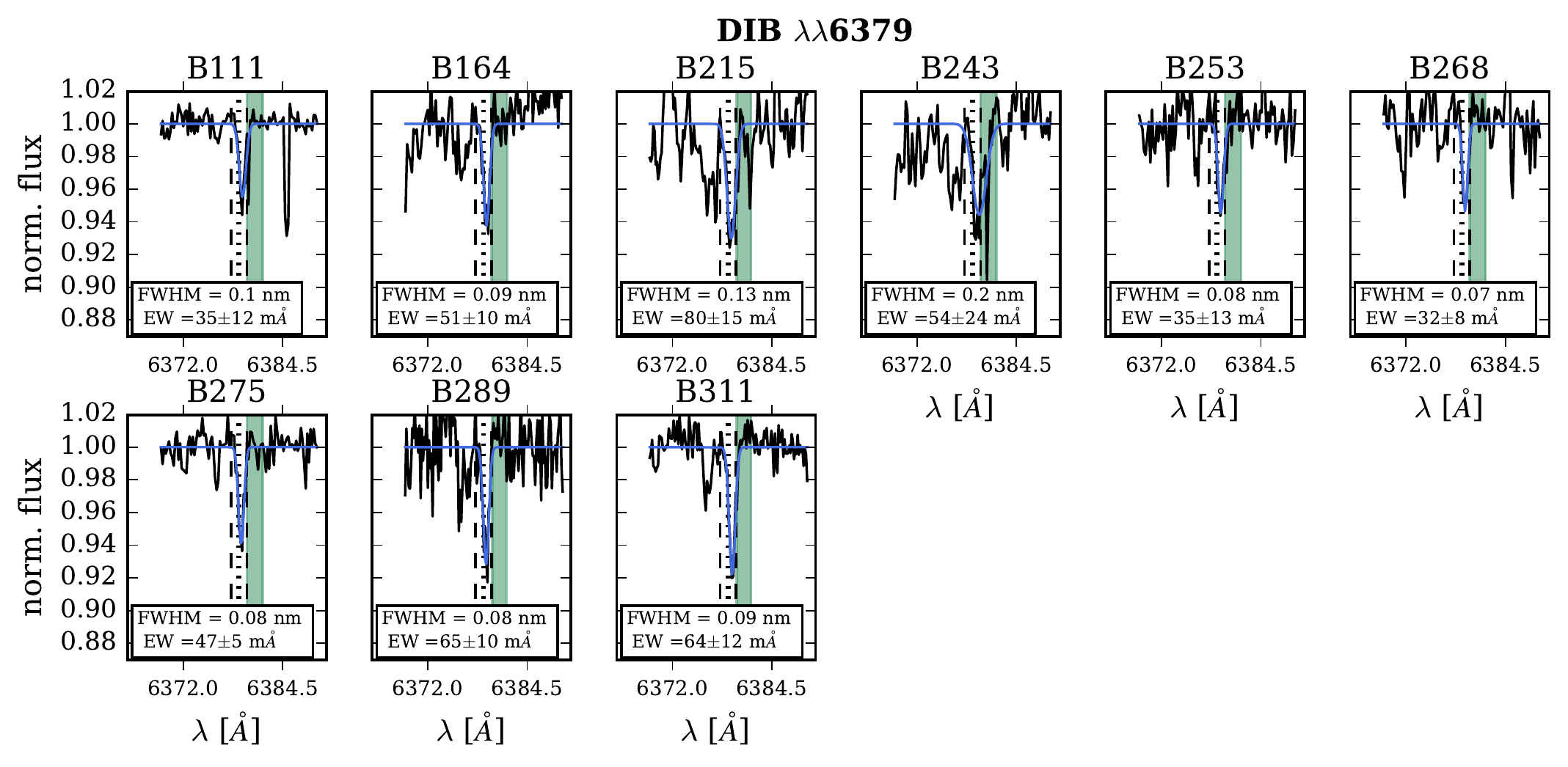}
    \caption{DIB profiles at 6379~\AA\ . The vertical dashed lines show the integration limits used to calculate the DIB strength, the dotted line shows the central wavelength of the DIB, and the green shaded region shows the region in which the error was calculated. With a solid-blue line we show the Gaussian fit to the DIB profile. The white box in the bottom of each panel indicated the FWHM and EW of this DIB for each object.}
    \label{P3:fig:ap6379}
\end{figure*}

% ______________8_______________

\begin{figure*}
    \centering
    \includegraphics[width=\hsize]{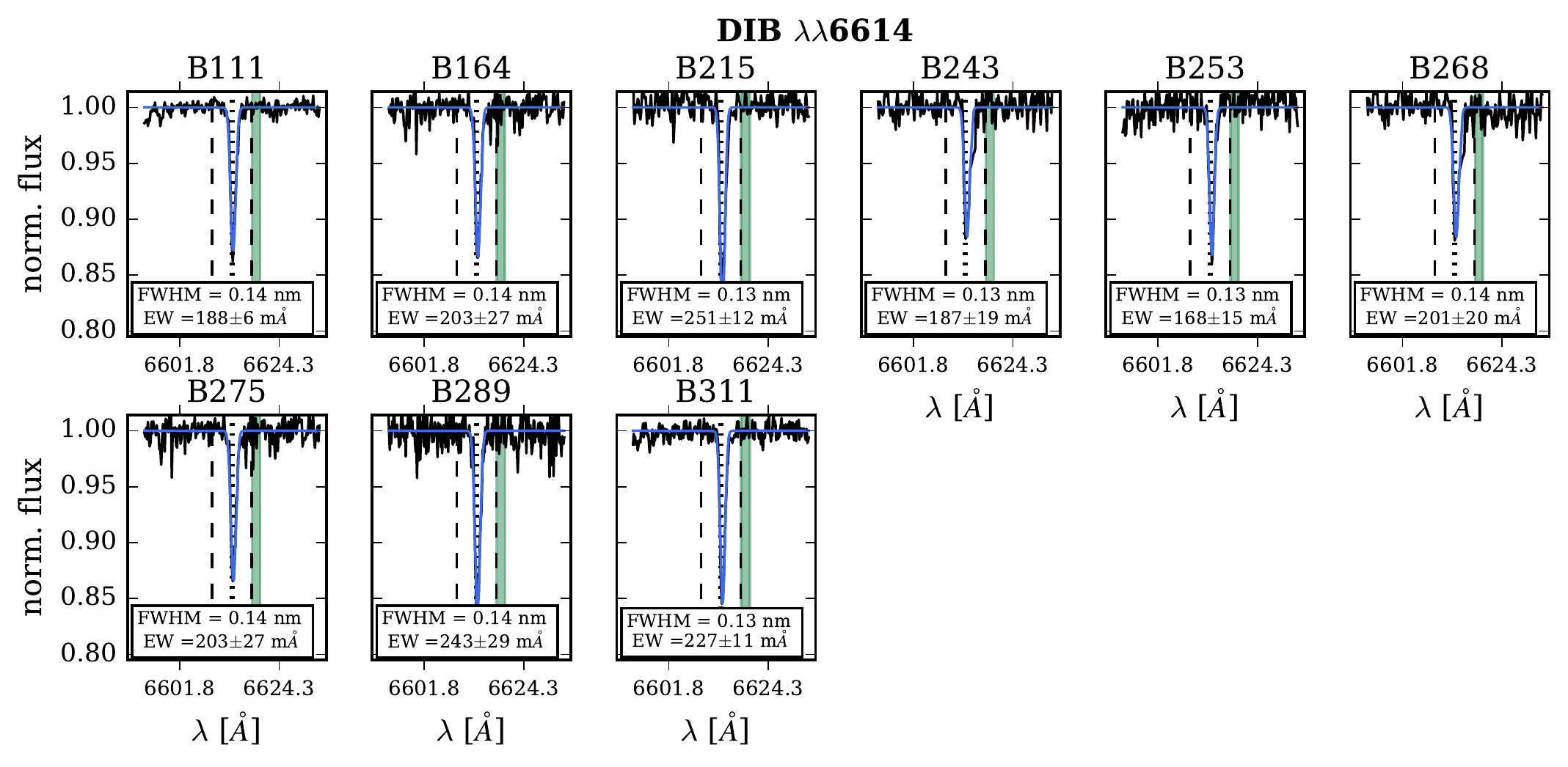}
    \caption{DIB profiles at 6113~\AA\ . The vertical dashed lines show the integration limits used to calculate the DIB strength, the dotted line shows the central wavelength of the DIB, and the green shaded region shows the region in which the error was calculated. With a solid-blue line we show the Gaussian fit to the DIB profile. The white box in the bottom of each panel indicated the FWHM and EW of this DIB for each object.}
    \label{P3:fig:ap6613}
\end{figure*}

% ______________8_______________

\begin{figure*}
    \centering
    \includegraphics[width=\hsize]{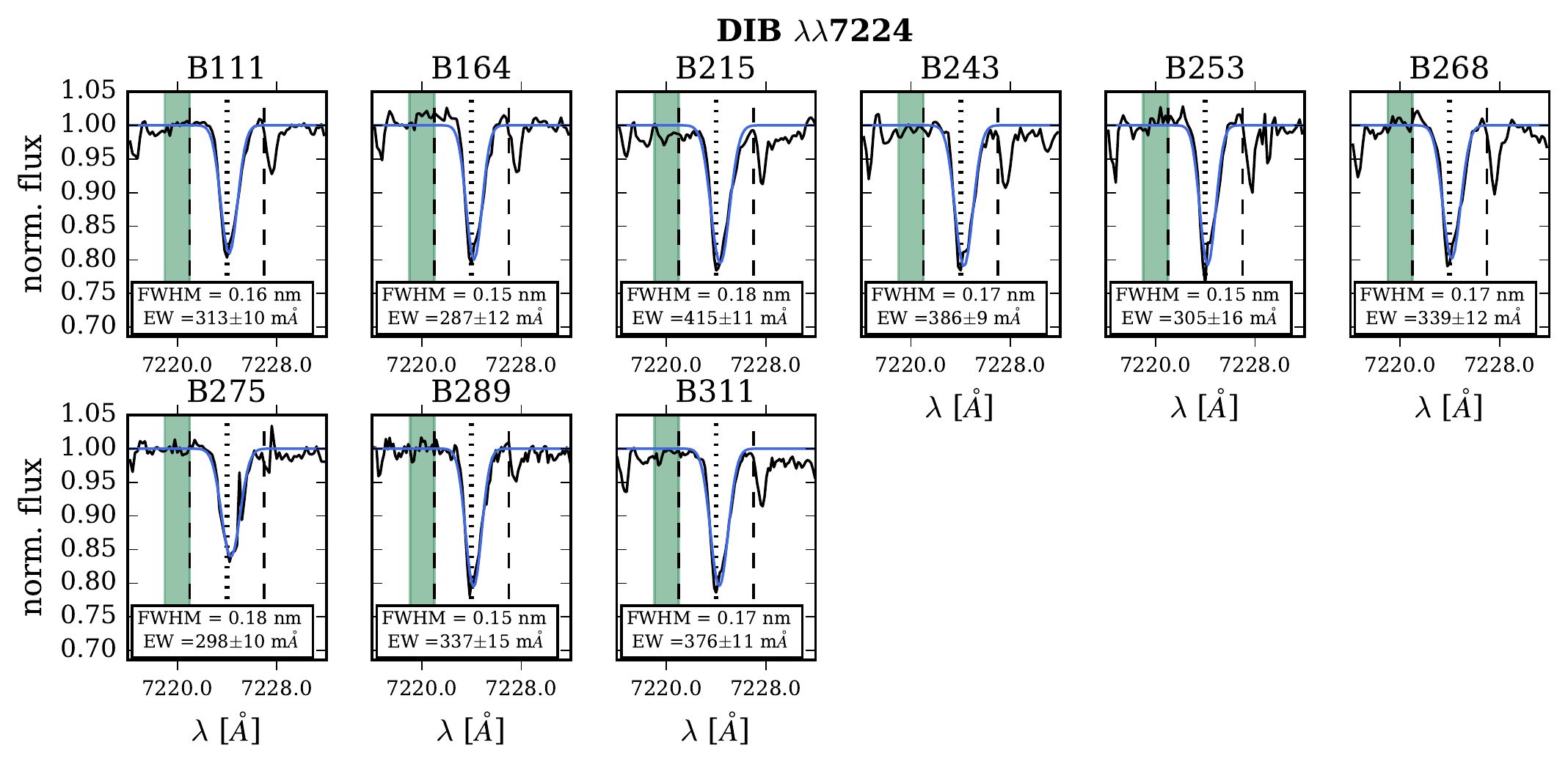}
    \caption{DIB profiles at 7224~\AA\ . The vertical dashed lines show the integration limits used to calculate the DIB strength, the dotted line shows the central wavelength of the DIB, and the green shaded region shows the region in which the error was calculated. With a solid-blue line we show the Gaussian fit to the DIB profile. The white box in the bottom of each panel indicated the FWHM and EW of this DIB for each object.}
    \label{P3:fig:ap7224}
\end{figure*}

% ______________8_______________

\begin{figure*}
    \centering
    \includegraphics[width=\hsize]{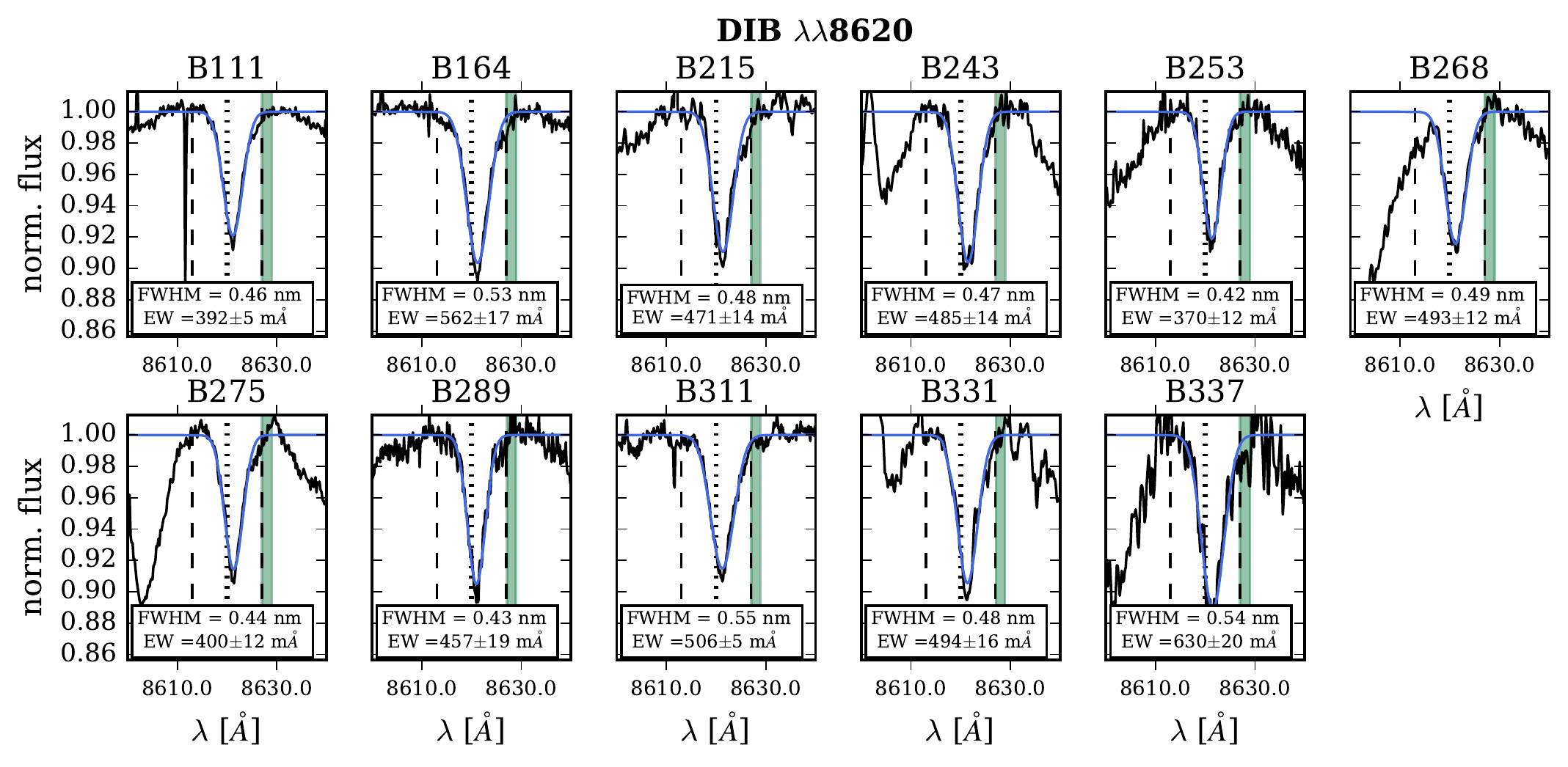}
    \caption{DIB profiles at 8620~\AA\ . The vertical dashed lines show the integration limits used to calculate the DIB strength, the dotted line shows the central wavelength of the DIB, and the green shaded region shows the region in which the error was calculated. With a solid-blue line we show the Gaussian fit to the DIB profile. The white box in the bottom of each panel indicated the FWHM and EW of this DIB for each object.}
    \label{P3:fig:ap8620}
\end{figure*}

% ______________9_______________

\begin{figure*}
    \centering
    \includegraphics[width=\hsize]{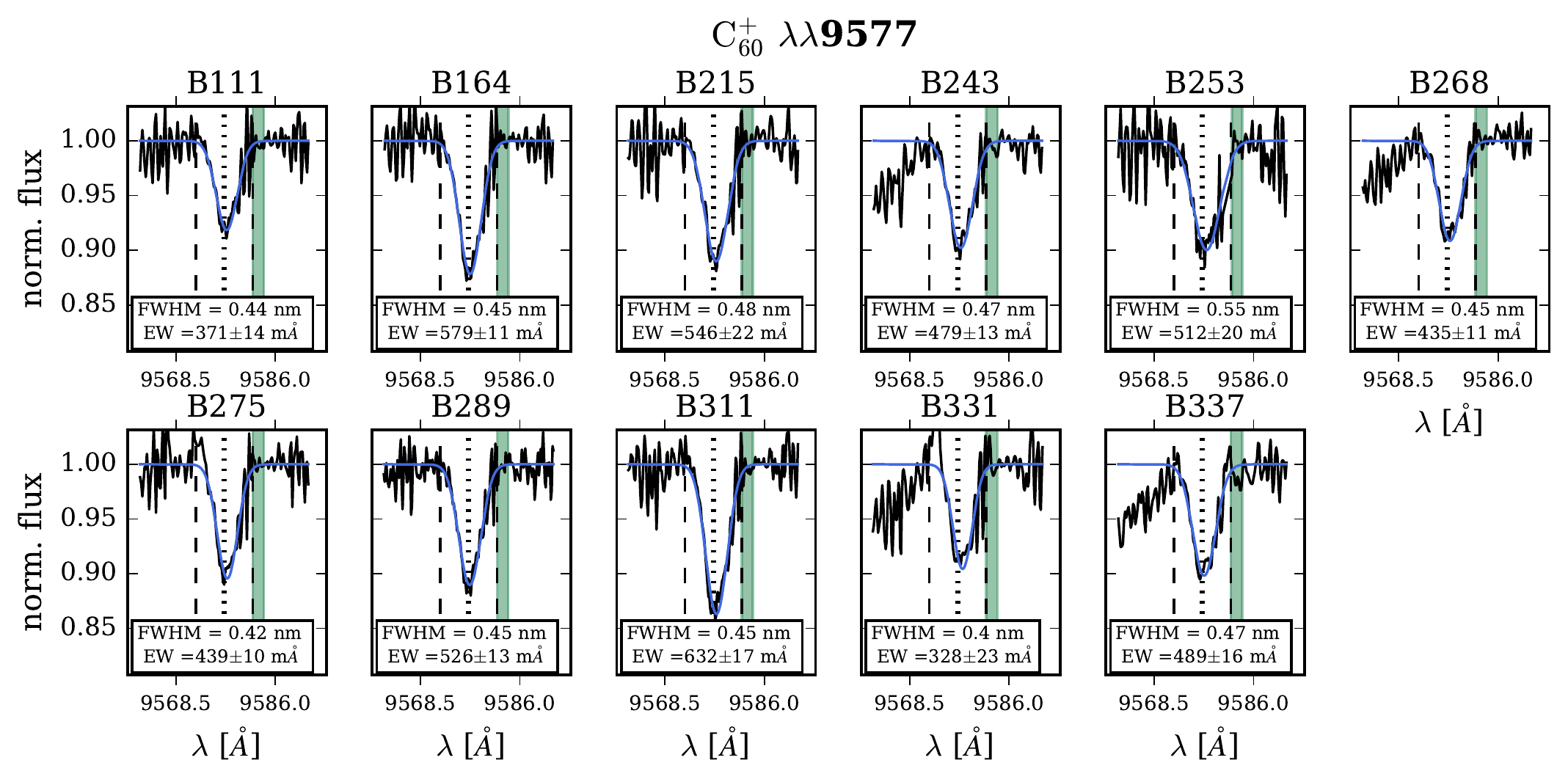}
    \caption{C$_{60}^+$ DIB profiles at 9577~\AA\ . The vertical dashed lines show the integration limits used to calculate the DIB strength, the dotted line shows the central wavelength of the DIB, and the green shaded region shows the region in which the error was calculated. With a solid-blue line we show the Gaussian fit to the DIB profile. The white box in the bottom of each panel indicated the FWHM and EW of this DIB for each object.}
    \label{P3:fig:ap9577}
\end{figure*}

% ______________10_______________

\begin{figure*}
    \centering
    \includegraphics[width=\hsize]{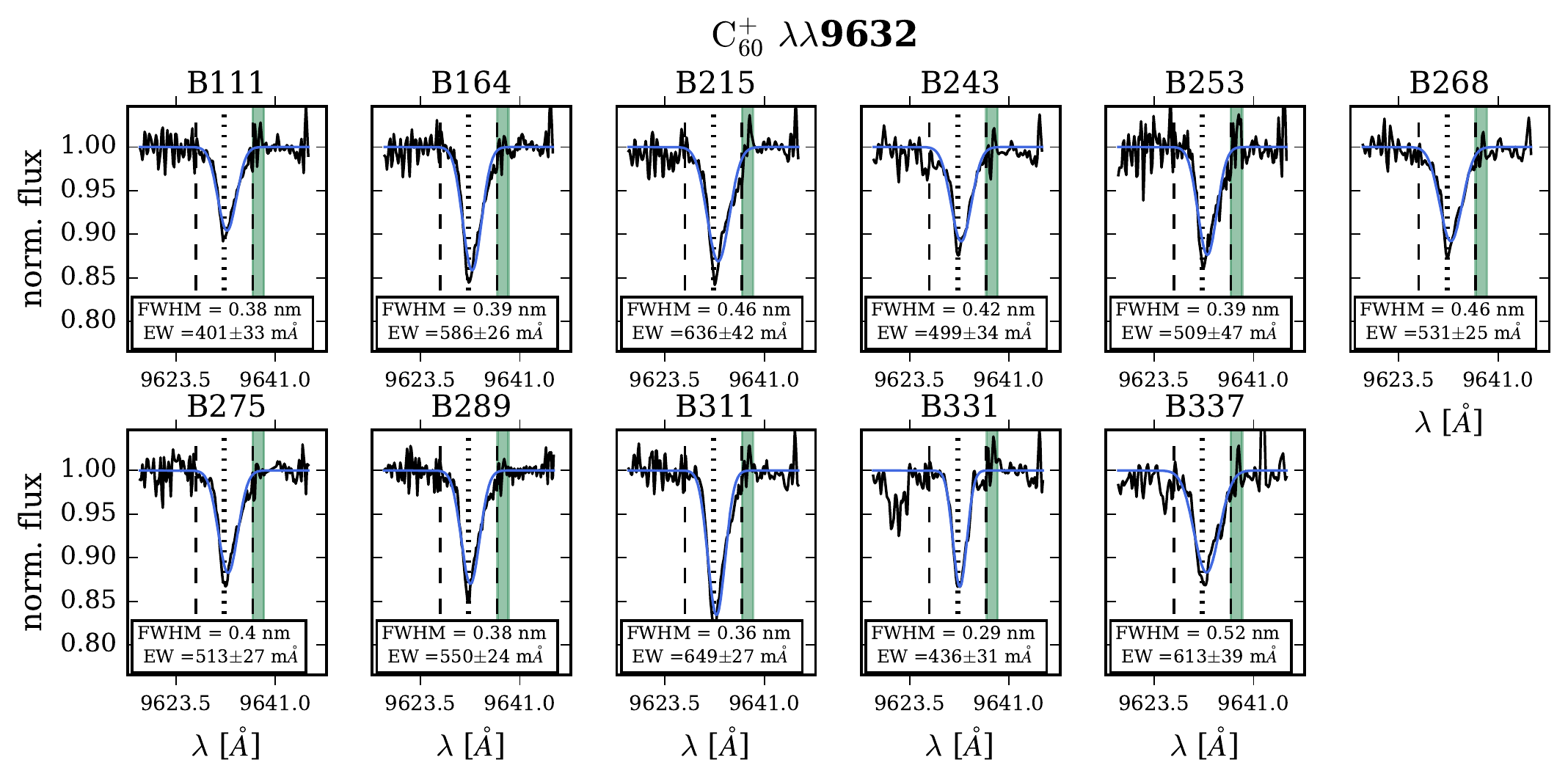}
    \caption{C$_{60}^+$ DIB profiles at 9632~\AA\ . The vertical dashed lines show the integration limits used to calculate the DIB strength, the dotted line shows the central wavelength of the DIB, and the green shaded region shows the region in which the error was calculated. With a solid-blue line we show the Gaussian fit to the DIB profile. The white box in the bottom of each panel indicated the FWHM and EW of this DIB for each object.}
    \label{P3:fig:ap963.2}
\end{figure*}

% ______________11_______________

\begin{figure*}
    \centering
    \includegraphics[width=\hsize]{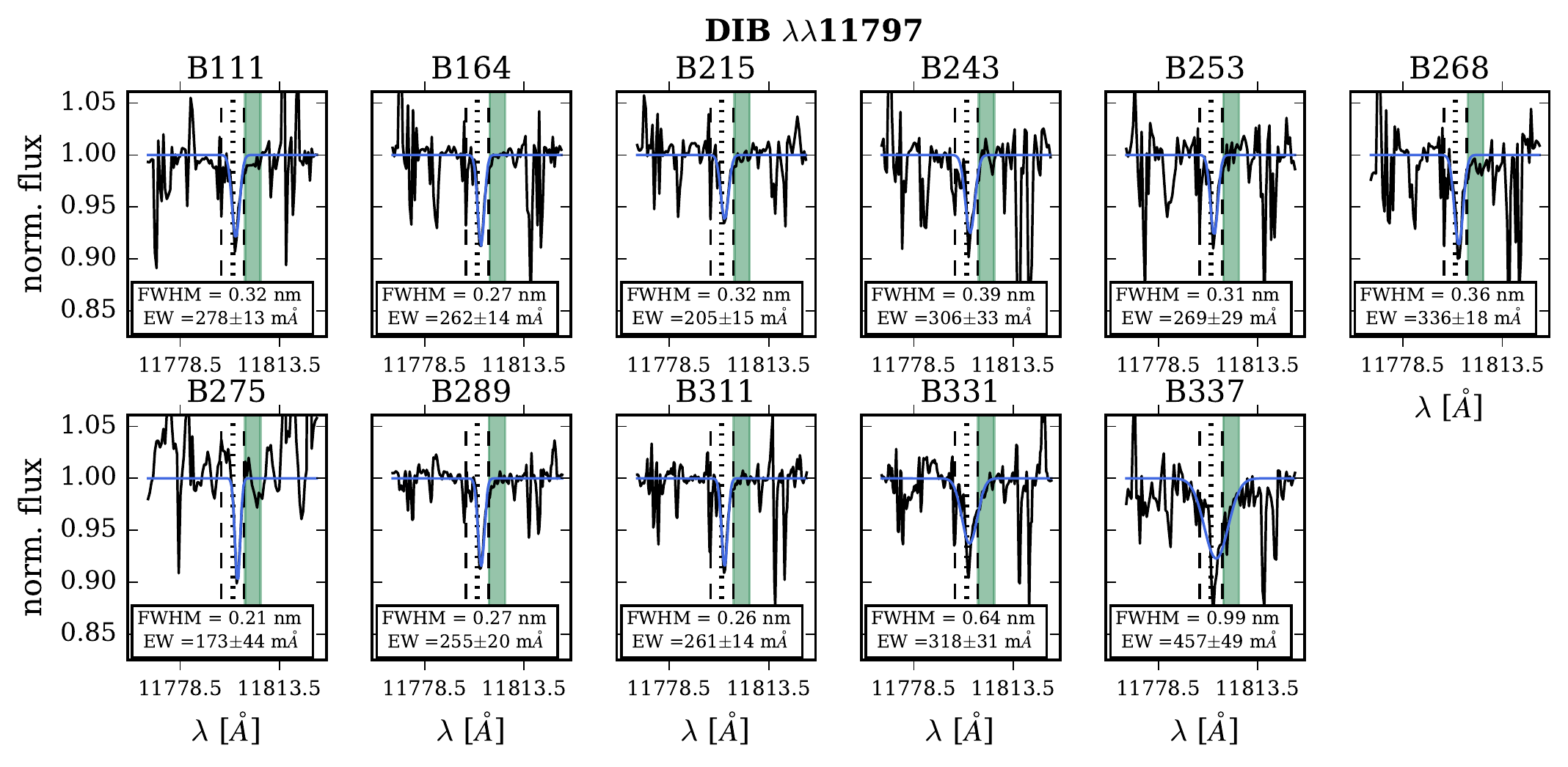}
    \caption{NIR DIB profiles at 11797~\AA\ . The vertical dashed lines show the integration limits used to calculate the DIB strength, the dotted line shows the central wavelength of the DIB, and the green shaded region shows the region in which the error was calculated. With a solid-blue line we show the Gaussian fit to the DIB profile. The white box in the bottom of each panel indicated the FWHM and EW of this DIB for each object.}
    \label{P3:fig:ap11797}
\end{figure*}

% ______________12_______________

\begin{figure*}
    \centering
    \includegraphics[width=\hsize]{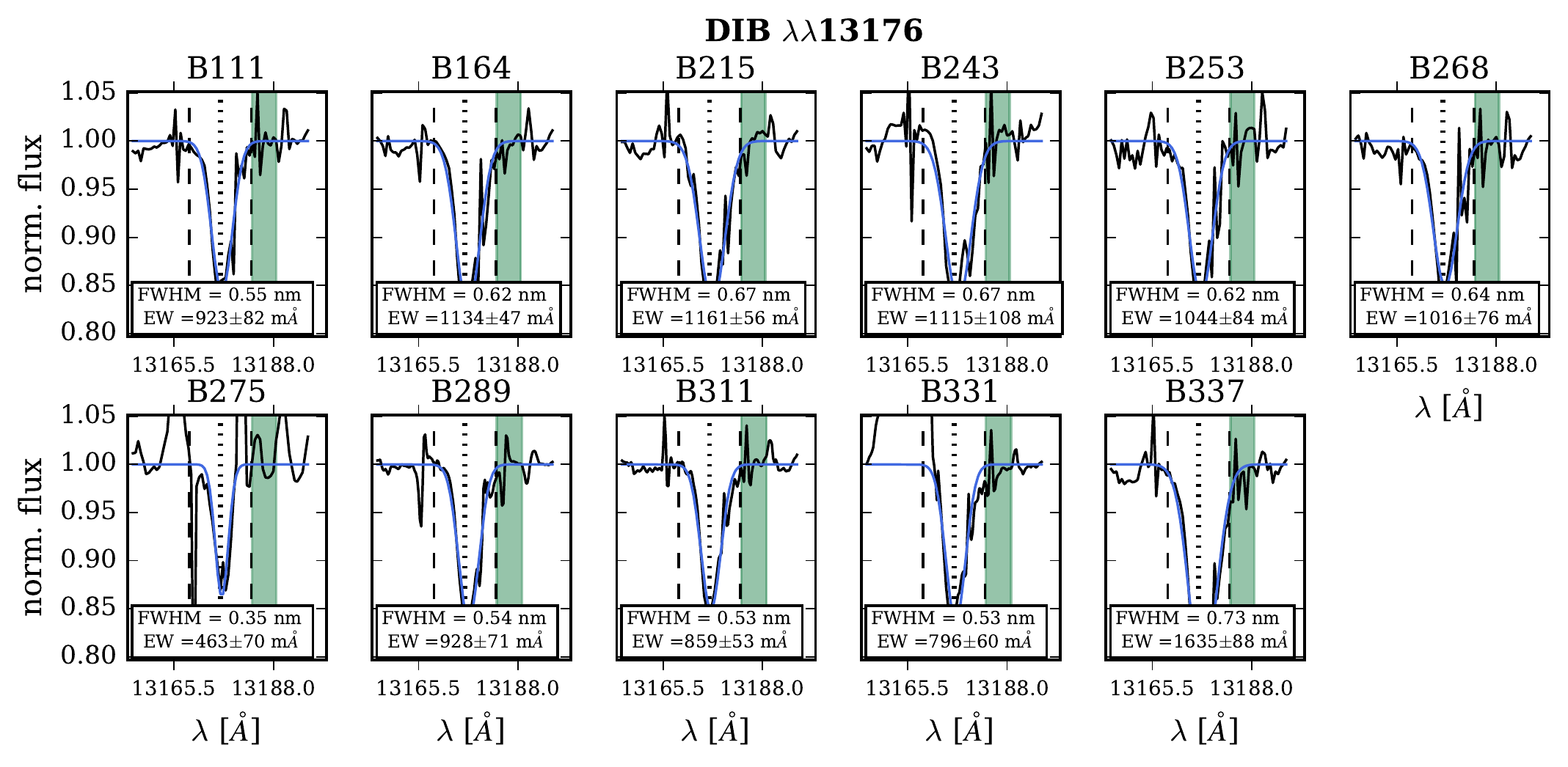}
    \caption{NIR DIB profiles at 13176~\AA\ . The vertical dashed lines show the integration limits used to calculate the DIB strength, the dotted line shows the central wavelength of the DIB, and the green shaded region shows the region in which the error was calculated. With a solid-blue line we show the Gaussian fit to the DIB profile. The white box in the bottom of each panel indicated the FWHM and EW of this DIB for each object.}
    \label{P3:fig:ap13176}
\end{figure*}

% ______________13_______________

\begin{figure*}
    \centering
    \includegraphics[width=\hsize]{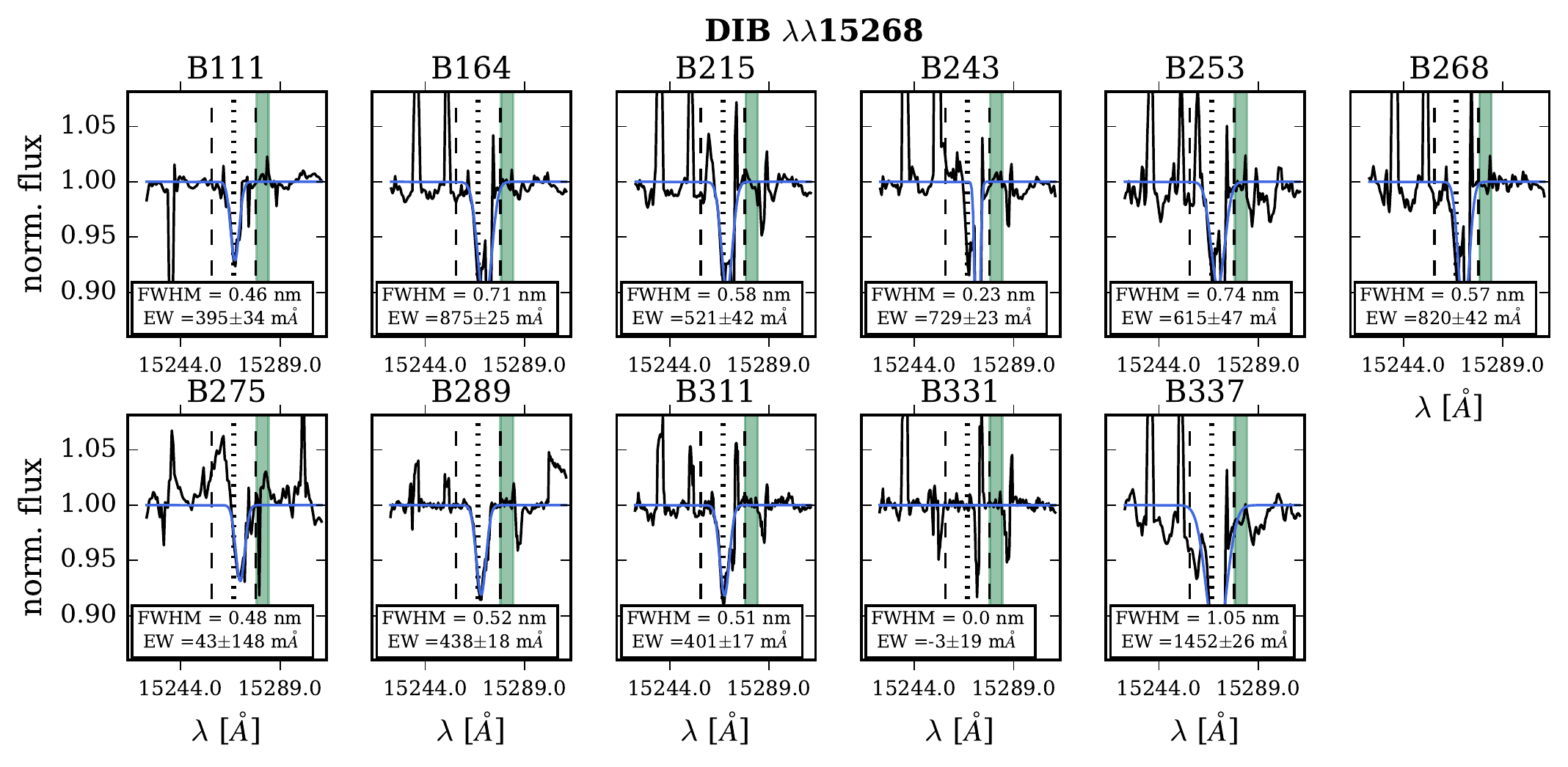}
    \caption{NIR DIB profiles at 15268~\AA\ . The vertical dashed lines show the integration limits used to calculate the DIB strength, the dotted line shows the central wavelength of the DIB, and the green shaded region shows the region in which the error was calculated. With a solid-blue line we show the Gaussian fit to the DIB profile. The white box in the bottom of each panel indicated the FWHM and EW of this DIB for each object.}
    \label{P3:fig:ap15268}
\end{figure*}

\end{appendix}

\end{document}